\documentclass[aps,prl,superscriptaddress, twocolumn]{revtex4-1}
\usepackage{graphicx}% Include figure files
\usepackage{dcolumn}% Align table columns on decimal point
\usepackage{bm}% bold math
\usepackage{amsmath}

\usepackage[mathlines]{lineno}% Enable numbering of text and display math
\usepackage{color}
\usepackage{bm,graphicx,hyperref}
\hypersetup{%
  breaklinks = {true},
  citecolor = {blue},
  colorlinks = {true},
  linkcolor = {red},
}

\newcommand{\mst}{\ensuremath{m_{\rm st}}}

\def\*#1{\mathbf{#1}}

\usepackage{setspace}
\begin{document}

\title{Anomalous thermal expansion in Ising-like puckered sheets}

\author{Paul~Z.~Hanakata}
\affiliation{Department of Physics, Harvard University, Cambridge, Massachusetts 02138, USA}
\email{paul.hanakata@gmail.com}

\author{Abigail Plummer}
\affiliation{Department of Physics, Harvard University, Cambridge, Massachusetts 02138, USA}
%\email{paul.hanakata@gmail.com}

\author{David~R.~Nelson}
\affiliation{Department of Physics, Harvard University, Cambridge, Massachusetts 02138, USA}
\email{drnelson@fas.harvard.edu}

\date{\today}
%\linenumbers
\begin{abstract}
  Motivated by efforts to create thin nanoscale metamaterials and
  understand atomically thin binary monolayers, we study the finite
  temperature statistical mechanics of arrays of bistable buckled
  dilations embedded in free-standing two-dimensional crystalline
  membranes that are allowed to fluctuate in three dimensions. The
  buckled nodes behave like discrete, but highly compressible, Ising
  spins, leading to a phase transition at $T_c$ with singularities in
  the staggered ``magnetization,'' susceptibility, and specific heat,
  studied via molecular dynamics simulations. Unlike conventional
  Ising models, we observe a striking divergence and sign change of
  the coefficient of thermal expansion near $T_c$ caused by the
  coupling of flexural phonons to the buckled spin texture. We argue
  that a phenomenological model coupling Ising degrees of freedom to
  the flexural phonons in a thin elastic sheet can explain this
  unusual response.
\end{abstract}

\maketitle
In recent decades, metamaterials with unique properties, such as
auxetic behavior and extreme stretchability, have been realized at the
macroscale~\cite{florijn-PRL-113-175503-2014, dias-sm-48-9087-2017} as
well as the nanoscale ~\cite{blees-Nature-524-204-2015,
  hanakata-PRL-121-255304-2018, hanakata-PRR-2-042006-2020,
  chen2020kirigami, bircan-NL-20-4850-2020}. More recently, there has
been growing interest in designing mechanical materials with
programmable memory, using multistable buckled
materials~\cite{florijn-PRL-113-175503-2014,
  oppenheimer-PRE-92-052401-2015,
  coulais-combinatorial-nature-535-529-2016,
  faber-advancedScience-7-2001955-2020, plummer-PRE-102-033002-2020,
  liu2021frustrating} and origami~\cite{silverberg-nature-14-389-2015,
  waitukaitis-PRL-114-055503-2015, overvelde-NatCom-7-1-2016,
  stern-PRX-7-041070-2017}.%, in which the functionality can be tuned by changing the ``states'' (configurations) as in a computer.

Buckled configurations have also been found (via either first-principles
simulations or experiments) in atomically thin materials such as
stanene, SnO, PbS, and borophane polymorphs~\cite{molle2017buckled,
  seixas-PRL-116-206803-2016, singh2017soft, pacheo-SnO-PRB-99-104108,
  hanakata-PRB-96-161401-2017, li2021-borophane, daeneke2017wafer},
as well as in graphene with topological defects or substitutional
impurities~\cite{lusk-PRL-100-175503-2008, zhang-JMPS-67-2-2014,
  grosso2015bending, hofer2019direct}.  At the nanoscale, thermal
fluctuations can strongly influence any mechanical memories stored in
the material as the energy barriers between bistable states become
comparable to the temperature. Furthermore, thermal fluctuations also
profoundly change the mechanics of atomically thin materials at long
wavelengths~\cite{gao-JMPS-66-42-2014, hanakata-EML-44-101270-2020,
  morshedifard-JMPS-149-104296-2021, kudin2001c,
  blees-Nature-524-204-2015, nicholl-NatCom-6-1-2015,
  nicholl-PRL-118-266101-2017}. Yet, few studies exist on the
interplay between microstructure (e.g., defects) and thermal
fluctuations in these atomically thin materials.
%Simulations and experiments have revealed scale-dependent
%elastic parameters, with thermal fluctuations increasing the bending
%rigidity and decreasing in-plane stiffness

We study here the thermal response and phase transitions of puckered
sheets with square arrays of buckled positive and negative dilational
defects using molecular dynamics simulations. We find that puckered
membranes undergo highly compressible Ising-like phase transitions. We
also observe an anomalous thermal expansion, where the typically
negative coefficient of thermal expansion briefly becomes positive
close to the transition, which we explain with a theoretical model
coupling spin and elastic degrees of freedom. Creating a highly
tunable coefficient of thermal expansion, spanning both positive and
negative values, is a goal of many metamaterial design efforts, and we
are not aware of any other physically realizable 2D material expected
to have this property \cite{burtch2019negative, boatti2017origami,
  greve2010pronounced}. This unusual anomaly could, for example, be
leveraged to create nanoscale device components whose dimensions are
insensitive to thermal changes at a particular operating temperature.
%section{The model}
%\includegraphics[width=0.7\textwidth]{Figures/schematics.pdf}
\begin{figure}
\begin{center}
\includegraphics[width=8.6cm]{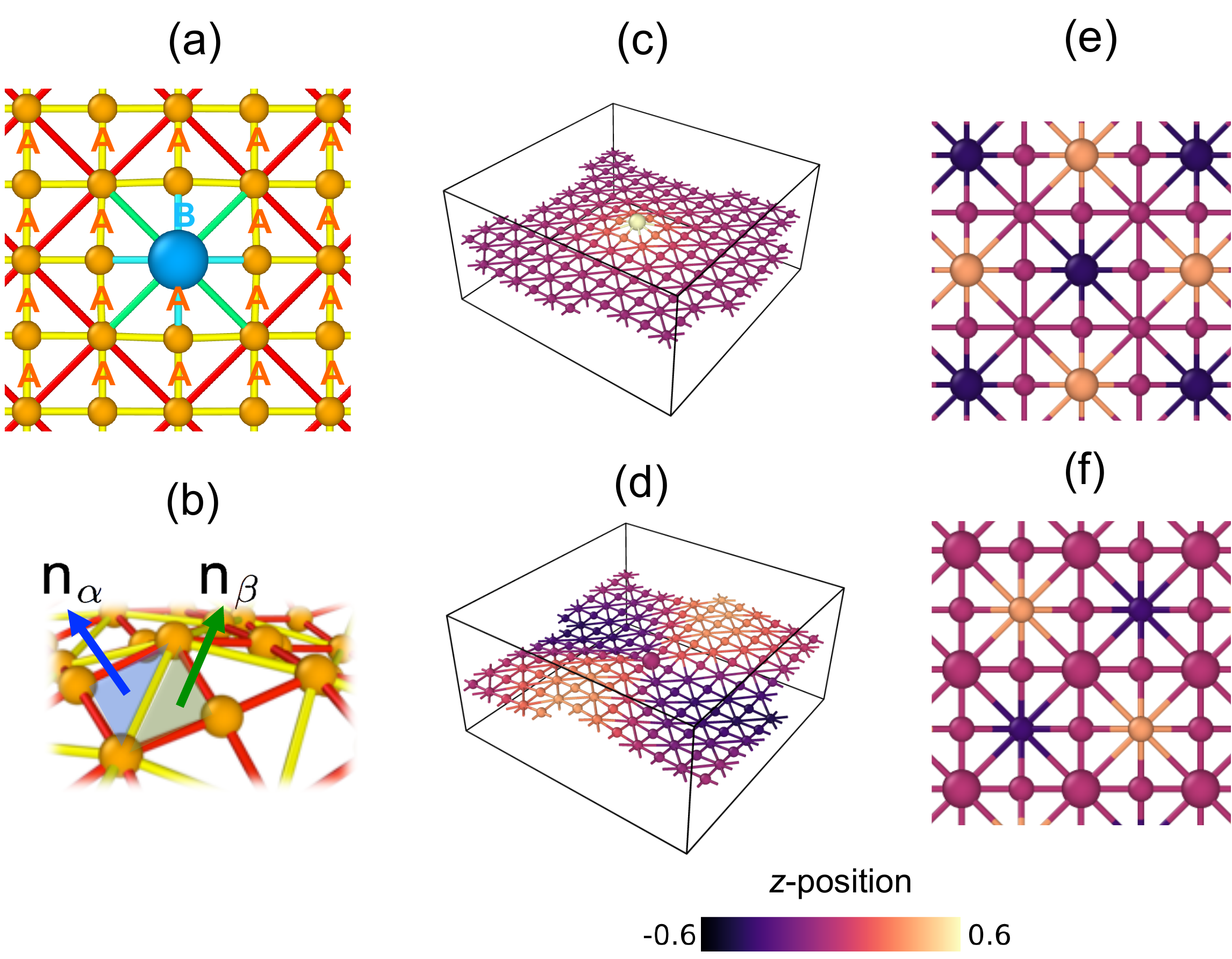}
\end{center}
\caption{(a) Square lattice model with background sites $\mathcal{A}$
  and a single dilation site $\mathcal{B}$. (b) Schematic of normals
  of two neighboring triangular plaquettes $\alpha, \beta$. (c,d)
  Height profiles of relaxed membranes with a single (c) positive and
  (d) negative dilation at $T=0$. The color represents the height
  relative to the zero plane in units of the lattice spacing
  $a_0$. The dilation nodes are indicated with a larger radius sphere.
  (e,f) Top views of membranes with a square array of positive (e) and
  negative (f) dilations in a (0, 2) array at $T=0$. Both display a
  checkerboard configuration characteristic of antiferromagnetism at
  $T=0$ when spins are defined as the nodes that buckle out of
  plane. Node positions are visualized using OVITO
  software~\cite{ovito}.}
  %  The membrane with a positive dilation (c) has a peaked profile whereas the membrane with a negative dilation (d) undergoes a saddle-like deformation.
  %  In terms of these spin variables, the buckled dilations in (e) are in a checkerboard configuration characteristic of antiferromagnetism at $T=0$. For the case of negative dilations, $\mathcal{A}$ nodes connected to four negative dilations buckle out of the plane, rather than the dilations themselves. We again observe a checkerboard configuration in (f). 
\label{fig:schematics}
\end{figure}

\emph{The model.---} Since {\it ab initio} molecular
dynamics~\cite{mehboudi2016structural} are computationally expensive
for studying phase transitions and atomistic potentials for puckered
materials are not yet developed, we use a coarse-grained discrete
membrane model~\cite{seung-PRA-38-1005-1988}, tuned to approximate an
isotropic elastic sheet in the continuum limit. Nodes are connected by
harmonic springs (Fig. \ref{fig:schematics}(a)) and there is an
energetic cost when the normals of neighboring planes are not aligned
(Fig. \ref{fig:schematics}(b)). The total energy, adapted from
ref. \cite{seung-PRA-38-1005-1988}, is given by
\begin{equation}
  E=\frac{k}{2}\sum_{\langle i,j\rangle}(|\pmb{r}_{i}-\pmb{r}_j|-a_{ij})^2+\hat{\kappa}\sum_{\langle \alpha, \beta \rangle}(1-\pmb{n}_{\alpha}\cdot\pmb{n}_{\beta}),
\end{equation}
where $k$ is the spring constant, $\hat{\kappa}$ is the microscopic
bending rigidity, and $a_{ij}$ is the rest length between two
connected nodes, $i$ and $j$. The first sum is over connected nodes
and the second sum is over neighboring triangular planes.  This model
(with $a_{ij}=a_0$, nodes on a triangular lattice) has been shown to
produce mechanical~\cite{zhang-JMPS-67-2-2014} and thermal
properties~\cite{hanakata-EML-44-101270-2020,
  morshedifard-JMPS-149-104296-2021} as well as height-height
correlation functions~\cite{bowick-PRB-95-104109-2017} consistent with
simulation results of 2D materials (e.g., graphene and MoS$_2$) using
atomistic potentials~\cite{zhang-JMPS-67-2-2014,
  roldan-PRB-83-174104-2011, jiang2014buckling} (Supplemental Material
(SM) Sec. VII)~\footnote{See Supplemental Material for details
  on theoretical derivations, extracting critical exponents,
  simulations and a supplementary video, which includes
  Refs.~\cite{koiter2009wt, kosmrljsphere, paulose, fire-algo,
    katsnelson2012graphene, stuart2000reactive,
    plimptonLAMMPS}}. Furthermore, anticipating our observations of
critical phenomena, we expect aspects of the behavior of the system to
be insensitive to microscopic details due to universality.

In this work, the nodes are arranged on a square lattice and the
dilation sites $\mathcal{B}$ are embedded into a background matrix of
standard, undilated sites $\mathcal{A}$. The dilations are modeled by
changing the preferred lengths of the bonds between $\mathcal{A}$ and
$\mathcal{B}$ sites~\cite{plummer-PRE-102-033002-2020}, mimicking
buckled monolayers (e.g., SnO and PbS~\cite{seixas-PRL-116-206803-2016,
  hanakata-PRB-96-161401-2017}). The rest lengths are:
$a_{\mathcal{AA}}=a_0, \overline{a}_{\mathcal{AA}}=a_0\sqrt{2},
a_{\mathcal{AB}}=a_0(1+\epsilon),
\overline{a}_{\mathcal{AB}}=a_0\sqrt{2(1+\epsilon+\epsilon^2/2)}$,
where $\epsilon$ is the fractional change in the bond length, and
$\overline{a}$ denotes a diagonal bond.
%~\footnote{We note that half of
%  the lattice sites have four bonds, and half have eight bonds. All
%  dilation sites $\mathcal{B}$ have eight bonds to better approximate
% isotropic expansions and contractions, and standard sites
%  $\mathcal{A}$ can have either four or eight bonds with
%  neighbors.}.
For pristine membranes, the corresponding continuum Young's modulus is
$Y=4k/3$ and the continuum bending rigidity is
$\kappa=\hat{\kappa}$~\cite{plummer-PRE-102-033002-2020}. The
continuum size of the dilation is defined as $\Omega_0=4a_0^2\epsilon$
\cite{plummer-PRE-102-033002-2020}. We choose microscopic elastic
parameters $a_0=1, k=100, \hat{\kappa}=1, \epsilon=\pm 0.1$. Here we
study membranes with dilations that provide positive and negative
extra area ($\Omega_0>0$ and $\Omega_0<0$ respectively) with periodic
boundary conditions in $x$ and $y$ directions. See the SM Sec. IV for
details on other parameter choices. Related tethered membrane models
have been studied before~\cite{nelson-PRB-27-2902-1983,
  radzihovsky-PRA-44-3525-1991, nelson-EPL-16-79-1991,
  kantor1992glassy}, but with quenched random disorder rather than
regular defect arrays.

\emph{Mapping buckled structures to Ising spins.}---We
first describe the behavior of the model at $T=0$. As the cost of
stretching/size of the dilation increases, the system crosses a
buckling threshold, and a subset of the nodes will prefer to buckle
out of the plane. As shown in Fig. \ref{fig:schematics}(c) and (d),
the relaxed configurations of isolated buckled positive and negative
dilations differ. The positive dilations create localized, peaked
structures, and the negative dilations lead to saddle-like
deformations. This difference can be understood by considering the
angular deficit/surplus at the dilation vertex in the inextensible
limit---positive dilations have a local angular deficit (discrete
positive Gaussian curvature) and negative dilations have a local
angular surplus (discrete negative Gaussian curvature).

Despite these differences, we can assign Ising spin variables to
dense, square arrays of either positive or negative dilations. In
arrays of positive dilations at $T=0$, the dilations themselves buckle
out of the plane (Fig. \ref{fig:schematics}(e)). In arrays of negative
dilations, the dilations remain in a single plane at $T=0$, and sites
on the lattice dual to the dilation superlattice buckle
(Fig. \ref{fig:schematics}(f)). We assign a spin variable of $\pm 1$
to each buckled site depending on whether the dilation/dual site
buckles up or down. At finite temperature, we assign spins using
nodes' positions relative to the local planes formed by their
neighbors to account for thermal fluctuations. With this mapping, the
buckled structures shown in Figs. \ref{fig:schematics}(e) and (f) are
equivalent to checkerboard spin configurations, mechanical analogs of
a nearest-neighbor Ising antiferromagnet (AFM).  Our simulations
support the conclusion that the AFM state is the lowest energy state
for the buckled positive and negative dilation arrays that we
study. See the SM, Sec. III and V and
\cite{plummer-PRE-102-033002-2020} for further discussion of the
buckling threshold and the ground states of arrays.

\emph{Finite temperature simulations.---} As we are
interested in the interplay between microstructure and temperature, we
perform molecular dynamics (MD) simulations of both pristine membranes
and membranes with positive and negative dilation defects at finite
temperature using HOOMD~\cite{anderson2020hoomd}.  The membranes have
$L_N\times L_N$ nodes with $L_N$ ranging from 24 to 192. Systems with
$L_N^2$ nodes have $N_I=\frac{L_N^2}{4}$ dilations. Temperatures are
reported in units of the bending energy $(\hat{\kappa}=1)$.  See the
SM Sec. IV and V for more simulation details. 
%{\color{red} To assign up and down spins to buckled sites at finite
%  temperature, we use the nodes' positions relative to the local
% planes formed by their neighbors. This procedure accounts for
%  thermal fluctuations relative to the zero plane, which can be much
%  larger than the typical dilation buckling amplitude in this highly
% compressible medium (approximately $0.4 a_0$). }

\begin{figure}[htp]
\begin{center}
\includegraphics[width=8.6cm]{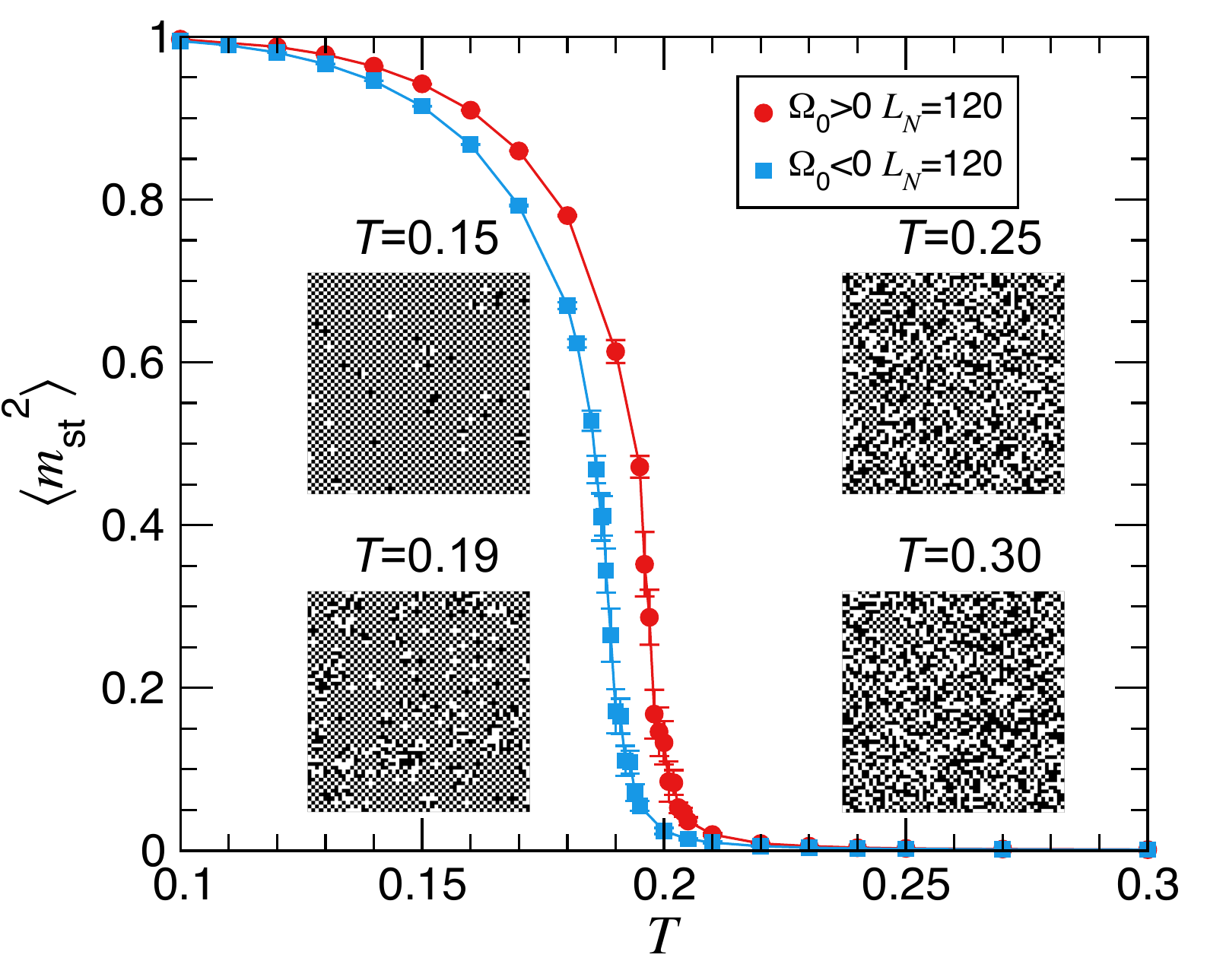}
\end{center}
\caption{Squared staggered magnetization $\langle \mst^2\rangle$ as a
  function of temperature $T$ for $L_N=120$. Plots for other system
  sizes can be found in the SM, Sec. VI. Error bars are calculated
  with between 10 and 50 runs, as described in the SM,
  Sec. IV. Jackknife method (see, e.g.,~\cite{Young2015}) is used to
  estimate statistical errors. The insets show snapshots of spin
  configurations of membranes with positive dilations ($\Omega_0>0$)
  for $T=0.15, 0.19, 0.25, 0.30$.  The spin configurations for
  membranes with negative dilations are similar.}
\label{fig:mag}                
\end{figure}

\begin{figure}[htp]
\begin{center}
\includegraphics[width=8.6cm]{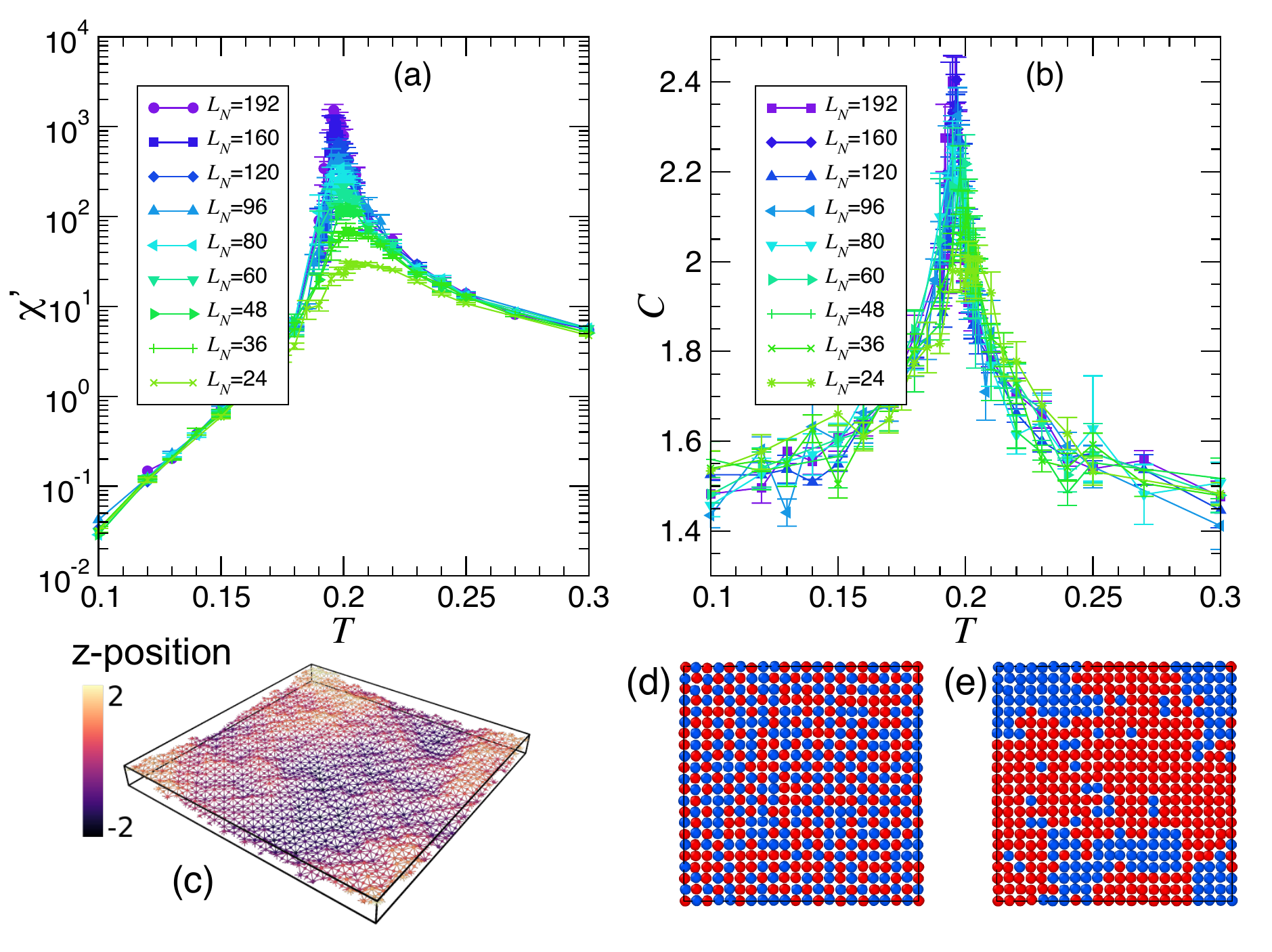}
\end{center}
\caption{(a) Staggered susceptibility $\chi^\prime$ and (b) specific
  heat $C$ as a function of temperature $T$ for different system sizes
  for membranes with positive dilations. Plots for peaks as a function
  of system size and plots for membranes with negative dilations can be
  found in the SM, Sec. VI. (c) Snapshot of a fluctuating puckered
  surface close to $T_c$. (d) Top view of up/down buckled sites
  (red/blue) and (e) the corresponding staggered spin configuration
  for the surface pictured in (c).}
\label{fig:Xi_and_C}
\end{figure}

\emph{Magnetic ordering and phase transitions.---} The mapping between
buckled structures and Ising spins suggests we can observe a
``magnetic" phase transition at finite temperature in our MD
simulations.  We use the staggered magnetization per spin as the order
parameter $\mst=\frac{1}{N_I}\sum_is_i(-1)^{x_i+y_i}$, where
$s_i=\pm1$ is the spin on site $i$, and $x_i, y_i$ are the site
indices on a 2D square lattice
(Fig. ~\ref{fig:Xi_and_C}d,e). Figure~\ref{fig:mag} shows
$\langle\mst^2\rangle$ for puckered membranes as a function of $T$.
We see clearly that pronounced AFM order for $T<0.2$ rapidly becomes
much smaller for $T>0.2$. Snapshots of spin configurations for several
temperatures are shown in Fig.~\ref{fig:mag} and the SM, Sec. V. Note
that in our model the bond topology remains unchanged across the
temperature range studied.

In studies of critical phenomena, it is typical to measure diverging
quantities such as the magnetic susceptibility $\chi$ and specific heat
$C$. 
%The quantities $\chi$ and $C$ can be calculated by measuring
%fluctuations in spins and energy, respectively. 
Following standard methods~\cite{binder-RepProgPhys-1997,
  landau-PRB-44-5081-1991, sandvik-AIP-2010}, we calculate the
staggered susceptibility as
$\chi^\prime= \frac{N_{\rm I}}{k_{\rm B}T}\left(\langle \mst^2\rangle
  -\langle |\mst|\rangle^2\right)$.
This computationally convenient quantity differs from the true
susceptibility by a constant factor above the transition and does not
affect the susceptibility exponents~\cite{binder-RepProgPhys-1997,
  landau-PRB-44-5081-1991}. See the SM, Sec. VI for details.
%We
%emphasize that for this calculation, we only consider fluctuations in
%the discrete pucker heights defined above.  
We also calculate the specific heat per site as
$C=\frac{1}{Nk_{\rm B}T^2}(\langle E^2\rangle-\langle E\rangle^2)$.
This measurement uses the \emph{total} potential energy, so $N$
includes all sites.
%not only the $N_{\rm I}$ spins.

The staggered susceptibility and specific heat of membranes with
positive dilations as a function of $T$ for a wide range of system
sizes are shown in Fig.~\ref{fig:Xi_and_C}. We see that $\chi^\prime$
and $C$ reach maxima at $T\simeq0.2$ and the peaks increase with
system size, a signature of phase transitions in finite
systems. Similar results for membranes with negative dilations appear
in the SM, Sec. VI. In finite systems, the correlation length
$\xi \sim |T-T_c|^{-\nu}$ cannot exceed the system size and thus the
diverging quantities will reach a maximum when $\xi\simeq L$. Finite
size scaling allows us to extract critical exponents
~\cite{binder-RepProgPhys-1997, landau-PRB-44-5081-1991,
  sandvik-AIP-2010}.
  
Upon fitting the data with power law functions, we measure
$\gamma/\nu=1.741\pm0.062, \alpha/\nu=0.068\pm0.018$ for $\Omega_0>0$
and $\gamma/\nu=1.684 \pm 0.061, \alpha/\nu=0.074\pm0.016$ for
$\Omega_0<0$. The measurement of $\gamma/\nu$ is consistent with the
rigid 2D Ising model value, $7/4$. The value of $\alpha/\nu$, on the
other hand, appears to be approximately four standard deviations away
from $0$, the 2D Ising expectation. Although our specific heat data
cannot completely exclude a rigid Ising model logarithmic divergence
in the specific heat, this observation suggests that the universality
class is not 2D Ising. We can plausibly attribute this departure to
the long-range interaction between staggered magnetization and
Gaussian curvature that arises in the phenomenological model
introduced in the following section. In the SM, Sec. VI, we extract
$\nu$ and present data for the exponents $\alpha$, $\gamma$, and
$\beta$.

%\section{Anomalous thermal expansion}
\begin{figure}[htp]
\begin{center}
\includegraphics[width=8.6cm]{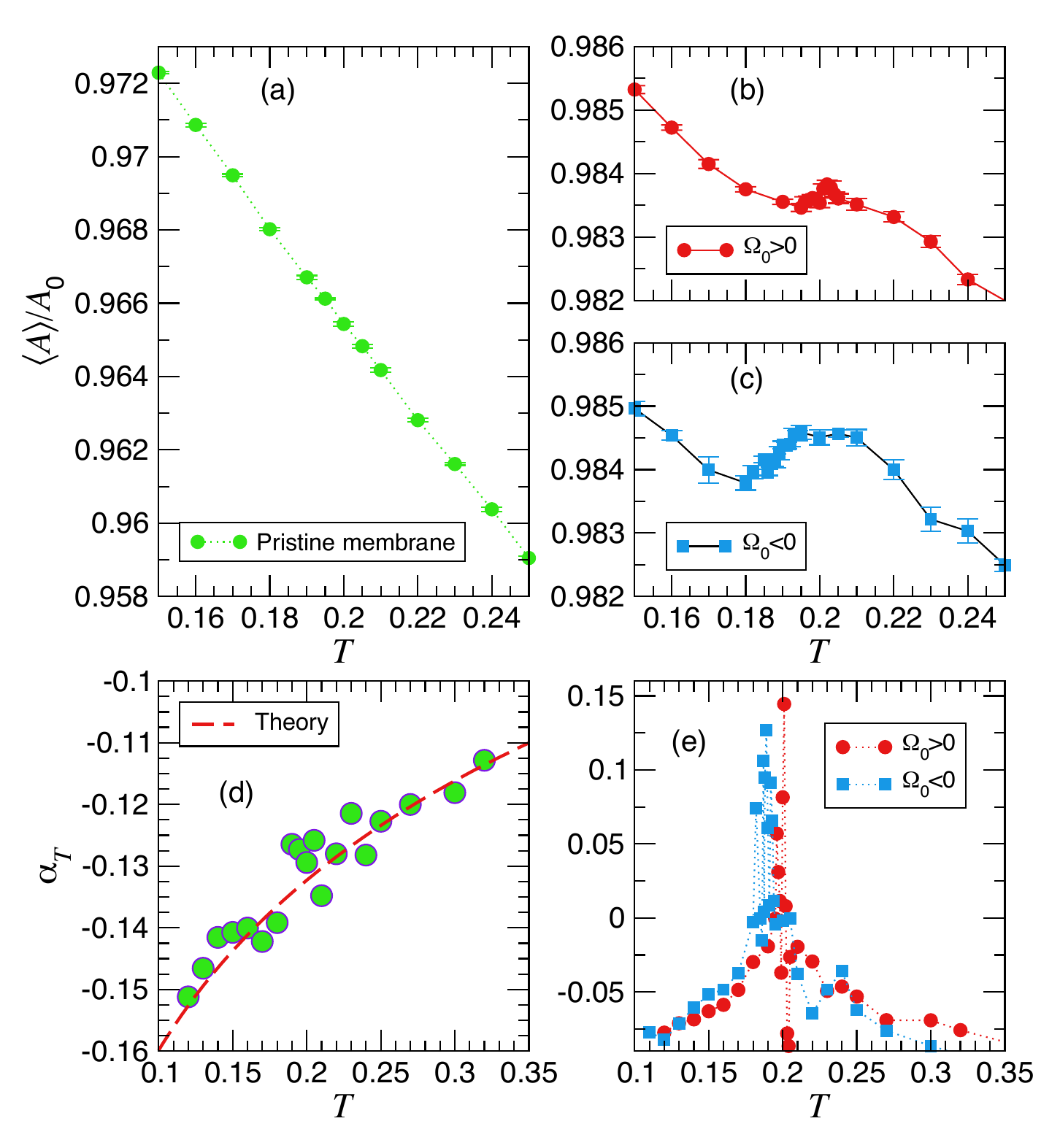}
\end{center}
\caption{Top row: Normalized area $\langle A\rangle/A_0$ as a function
  of $T$ for (a) pristine membranes, (b) membranes with positive
  dilations, and (c) membranes with negative dilations for
  $L_N=120$. $\langle A\rangle/A_0$ decreases with increasing $T$ for
  pristine membranes whereas $\langle A\rangle/A_0$ for puckered
  membranes shows non-monotonic behavior. Bottom row: The coefficient
  of thermal expansion $\alpha_T$ as a function of $T$ for (d)
  pristine membranes and (e) membranes with dilations. The theoretical
  prediction of $\alpha_T$ for pristine membranes with \emph{no}
  adjustable fitting parameters matches very well with simulations
  (red dashed line). Far below $T_c$, $\alpha_T$ for membranes with
  dilations is negative, as for pristine membranes. Close to $T_c$,
  $\alpha_T$ increases rapidly and reaches a \emph{positive} value,
  decreasing again to a negative value for $T$ well above $T_c$. }
\label{fig:alpha-T}
\end{figure}

\emph{Anomalous thermal expansion.---} The order-disorder transition
has a striking effect on the thermal expansion of the membrane as a
function of temperature. We first examine the thermal contraction of a
pristine membrane (no dilations) to establish a point of
comparison. Thermalized membranes have been studied extensively using
MD and Monte Carlo simulations ~\cite{roldan-PRB-83-174104-2011,
  gao-JMPS-66-42-2014, yllanes-NatCom-8-1-2017,
  bowick-PRB-95-104109-2017, wan2017thermal,
  morshedifard-JMPS-149-104296-2021, hanakata-EML-44-101270-2020}, and
their negative coefficient of thermal expansion $\alpha_T$ has been
calculated analytically ~\cite{kosmrlj-PRB-93-12-125431}.
\begin{equation}
\alpha_{T}=\frac{1}{A_0}\frac{dA}{dT}\simeq -\frac{k_{\rm B}}{4\pi\kappa}\left[  {\rm ln}\left(\frac{l_{\rm th}}{a_0}\right) + \frac{1}{\eta} -\frac{1}{2}   \right],
\label{eq:alpha}
\end{equation}
where the thermal length
$l_{\rm th}\equiv\frac{\pi}{q_{\rm
    th}}=\sqrt{\frac{16\pi^3\kappa^2}{3Yk_{\rm B}T}}$
and $\eta$ is a universal scaling exponent describing flexural
phonons, $\eta \approx 0.8$~\cite{kosmrlj-PRB-93-12-125431}. In our
simulations, we vary $T$ from 0.100 to 0.400, which varies
$l_{\rm th}$ from $\sim3.5a_0$ to $1.8a_0$.
Figure~\ref{fig:alpha-T}(a) shows the average projected area divided
by the area of a flat membrane as a function of $T$. Upon computing
$\alpha_T = \frac{1}{A_0} \frac{dA}{dT}$, we find excellent agreement
with Eq. \ref{eq:alpha} with no free parameters (red dashed line in
Fig. \ref{fig:alpha-T}(d)), using the zero-temperature values of the
bending rigidity and Young's modulus. The pristine membrane model
therefore reproduces the negative coefficient of thermal expansion of
materials such as graphene~\cite{yoon2011negative}. In contrast,
positive thermal expansion has been measured in relatively thick
freestanding transition metal dichalcogenides, possibly due to a
higher bending rigidity suppressing flexural phonons
~\cite{hu2018thermalexpansion,bertolazzi2011stretching,
  wang2019bending, jiang2014buckling}.

In contrast, $\langle A \rangle/A_0$ (and hence $\alpha_T$) for
puckered membranes shows non-monotonic behavior. Here, the constant
factor $A_0$ is the projected area of the lowest energy state at
$T=0$, a buckled checkerboard as described above. We observe that,
while there is shrinkage for $T<T_c$ as for a pristine membrane, the
value of $\alpha_T$ is less strongly negative. For $T \ll T_c$,
$\alpha^{\rm puckered}_T/\alpha^{\rm pristine}_T\sim0.5$, suggesting
that membranes with ordered puckers stiffen. This observation is
consistent with a theoretical argument based on
\cite{kosmrlj-PRE-88-012136-2013}, treating the buckled dilation
texture as a frozen background metric (SM, Sec. II).  The calculation
predicts the existence of an increased bending rigidity at $T=0$,
$\kappa_R\approx \kappa\left(1+\frac{3 Y h_0^2}{32 \kappa}\right)$,
where $h_0$ is the amplitude of the buckled membrane. Close to the
transition, however, $\alpha_T$ increases rapidly and eventually
reaches a \emph{positive} value. Evidently, the swelling due to
disordered up and down puckers on all length scales near $T_c$
dominates the entropic shrinkage present in pristine
sheets~\cite{kosmrlj-PRB-93-12-125431, yoon2011negative,
  hu2018thermalexpansion}.

\emph{Phenomenological model.---} To better understand the observed
differences between the thermal expansion of pristine membranes and
membranes with dilations, we introduce a ``flexural Ising model," with
an effective free energy that couples an Ising order parameter to a
thin elastic sheet that is allowed to fluctuate both in and out of the
plane. We assume coarse-graining such that the short wavelength,
impurity-scale phonons are accounted for by a staggered pucker order
parameter $\mst$, which interacts with a long wavelength nonlinear
strain matrix, $u_{ij}$.
\begin{align}
F= & \int d^2 x\bigg[\frac{\kappa}{2}  \left( \nabla^2 f\right)^2 + \mu u_{ij}^2 + \frac{\lambda}{2} u_{kk}^2+ \frac{K}{2} \left(\nabla \mst \right)^2\nonumber\\
&+\frac{r}{2} \mst^2+ u  \mst^4 + g \mst^2 u_{kk} \bigg],
\label{eq:freeen}
\end{align}
where $u_{ij}$ is related to in-plane displacements $u_j$ and
out-of-plane displacements $f$ by
$u_{ij}=\frac{1}{2}\left(\frac{\partial u_{i}}{\partial
    x_{j}}+\frac{\partial u_{j}}{\partial x_{i}}+\frac{\partial
    f}{\partial x_{i}} \frac{\partial f}{\partial x_{j}}\right)$
\cite{landau}.  The term proportional to $g$ is the lowest order
contribution allowed by symmetry coupling the phonon and order
parameter fields. Similar free energies have been used to study flat
compressible 2D Ising models in the limit
$f=0$~\cite{sak-PRB-10-3957-1974,
  larkin-spj-29-891-1969}. We also note similarities to
  free energies used to model electron-phonon interactions in
  graphene~\cite{gazit2009correlation, guinea2014collective,
    bonilla2016critical, cea2020numerical}.

Upon tracing out the in-plane phonons according to standard methods
\cite{nelsonpeliti, sak-PRB-10-3957-1974, larkin-spj-29-891-1969},
Eq. \ref{eq:freeen} becomes: \small
\begin{align}
F_{\text{eff}}=& \frac{g^2}{2A_0}\left( \frac{1}{2 \mu+\lambda}-\frac{1}{ \mu+\lambda}\right) \left(\int d^2 x \mst^2 \right)^2\nonumber\\
&+\int^\prime d^2 x \left[ \frac{Y}{8} \left(P_{ij}^T \partial_i f \partial_j f \right)^2 +  \frac{g \mu}{2 \mu+\lambda} \left( \mst^2 P_{ij}^T \partial_i f \partial_j f \right) \right]\nonumber \\
&+ \int d^2 x \bigg[ \frac{\kappa}{2} (\nabla^2 f)^2 + \frac{K}{2}\left(\nabla \mst\right)^{2}+\frac{r}{2} \mst^{2}\nonumber\\
&+\left(u- \frac{g^2}{2(2\mu+\lambda)}\right) \mst^{4} \bigg],
\label{eq:efffreeen}
\end{align}
\normalsize where $P_{ij}^T$ is the transverse projection operator and
the primed integral omits $\*q=0$ modes. Equation \ref{eq:efffreeen}
has three terms that are not present for either pristine membranes or
the Ising model. The first and final terms, proportional to $g^2$,
also appear for flat compressible 2D Ising models
\cite{sak-PRB-10-3957-1974, brunosak}. The term proportional to $g$,
however, is unique to the flexural Ising model. Since the Laplacian of
$-\frac{1}{2} P_{ij}^T \partial_i f \partial_j f$ is the Gaussian
curvature $S(\*x)$ in the Monge representation \cite{nelsonbook}, this
term represents a long-range interaction between the squared staggered
magnetization and the Gaussian curvature of the form
$\frac{1}{2\pi}\int d^2 x \int d^2 x^\prime \mst^2(\*x) S(\*x^\prime)
\ln(|\*x-\*x^\prime|)$.
A power counting argument suggests that the coefficient
$w \equiv \frac{g \mu}{2 \mu+\lambda}$ is a strongly relevant
operator. We plan to examine the behavior of $w$ in more detail in
future work.

We can calculate the coefficient of thermal expansion $\alpha_T$ by
adding an in-plane pressure, and compare to the simulation data in
Fig.~\ref{fig:alpha-T}. As shown in Sec. I of the SM, we find the
average change in area
\begin{equation}
\langle \delta A \rangle = - \frac{g A_0 \langle \mst^2 \rangle}{\mu+\lambda}- \frac{A_0}{2} \left\langle \left( \frac{\partial f}{\partial x_i}\right)^2 \right \rangle,
\end{equation}
and coefficient of thermal expansion
\begin{equation}
\alpha_T=\frac{1}{A_0} \frac{dA}{dT}=-\frac{d}{dT} \left(\frac{g \langle \mst^2 \rangle}{\mu+\lambda}\right) - \frac{d}{dT}\left\langle \frac{1}{2} \left(\frac{\partial f}{\partial x_i}\right)^2\right \rangle .
\end{equation}
We expect that the microscopic couplings $g$ and $\mu+\lambda$ depend
only weakly on temperature over the temperature range of interest,
provided we are far below the high temperature crumpling
transition. Therefore, the contribution from the first term is sharply
peaked around $T_c$, given the results in Fig~\ref{fig:mag}.  We
expect $g>0$, as the antiferromagnetic state has a smaller projected
area than the ferromagnetic state at $T=0$, consistent with the
positive peak in $\alpha_T$ at $T_c$ observed in
Fig.~\ref{fig:alpha-T}(e). The second term is the usual entropic
thermal shrinkage, also present for a pristine membrane
\cite{kosmrlj-PRB-93-12-125431}.

%\section{Discussion} 
\emph{Conclusion.---} We observe a phase transition in the staggered
magnetization of a puckered membrane, which provides a mechanical
analog of a highly compressible antiferromagnetic Ising model.
Furthermore, we find that the order-disorder transition produces an
anomalous thermal response for puckered membranes. These observations
suggest a strong coupling between flexural phonons and the ordering of
the spins (buckled sites). We introduce a phenomenological ``flexural
Ising model'' that anticipates a competing effect between entropic
shrinkage due to out-of-plane deformations and swelling due to pucker
disorder at the phase transition.

Our findings suggest that bistable buckled structures change the
thermal response of 2D materials, leading to a tunable coefficient of
thermal expansion. The ability to tune thermal expansion is important
for combining different materials, as mismatched thermal expansion can
affect the longevity of integrated
materials~\cite{werner2001review}. Materials with tunable thermal
expansion are rare and often require precise
engineering~\cite{burtch2019negative}.

Since the phase transition temperature in our model depends on the
elastic constants of the host lattice and the separation between
dilations, one could imagine constructing a nanocantilever or
nanoactuator out of a puckered membrane designed to be insensitive to
thermal expansion/shrinkage at the temperature at which it must work
($\alpha_T=0$ at two temperatures, one above and one below
$T_c$). Additionally, the mechanism of an inefficient packing of
buckled structures resulting in a global expansion can be applied to
macroscale materials with multistable units~\cite{boatti2017origami,
  faber-advancedScience-7-2001955-2020, liu2021frustrating}. Moreover,
our work suggests the possibility of studying novel universality
classes in 2D materials with a coupling between spin and both in-plane
and out-of-plane displacements, generalizing past work on compressible
Ising models to include flexural phonons.
\section{Acknowledgments}
PZH, AP, and DRN acknowledge support through the NSF grant DMR-1608501
and via the Harvard Materials Science Research and Engineering Center,
through NSF grant DMR-2011754. PZH, AP and DRN thank Suraj Shankar for
helpful discussions. PZH also thanks Adam Iaizzi for helpful comments.

%\section{Methods}

%
%\bibliography{biblio}
\end{document}

\title{Supplemental Material}

%\author{Paul~Z.~Hanakata}
%\affiliation{Department of Physics, Harvard University, Cambridge, Massachusetts 02138, USA}
%\email{paul.hanakata@gmail.com}
%\author{}
%\affiliation{Department of Physics, Harvard University, Cambridge, Massachusetts 02138, USA}
%\email{paul.hanakata@gmail.com}

%\author{Abigail Plummer}
%\affiliation{Department of Physics, Harvard University, Cambridge, Massachusetts 02138, USA}
%\email{paul.hanakata@gmail.com}

%\author{David~R.~Nelson}
%\affiliation{Department of Physics, Harvard University, Cambridge, Massachusetts 02138, USA}

%\date{\today}
%\linenumbers
%\begin{abstract}
%\end{abstract}
%\pacs{}
\maketitle
\vspace{-2em}
\tableofcontents

Here, we provide theoretical
derivations, as well as procedures for performing molecular dynamics simulations
and extracting critical exponents.
\section{Phenomenological model}\label{model}
We calculate here the ensemble-averaged area change
$\langle \delta A \rangle$ using our phenomenological model. To do so, we add an in-plane pressure term proportional to $\alpha$ to Eq. 3 of the main text
the form
\begin{equation}
-\alpha \int d^2x \left(\frac{\partial u_{k}}{\partial x_{k}}\right)\equiv-\alpha \int d^2x \tilde{u}_{kk}.
\label{eq:pressure}
\end{equation}
The change in the projected area is
\begin{align}
\langle \delta A \rangle &= \frac{\int \mathcal{D}
  \*u\int \mathcal{D}
 f \int \mathcal{D} \mst \left( \int d^2x \tilde{u}_{kk}\right) e^{-\beta F}}{\int \mathcal{D} \*u \int \mathcal{D}
 f \int \mathcal{D} \mst e^{-\beta F}}= \frac{\int \mathcal{D} \*u \int \mathcal{D}
 f \int \mathcal{D} \mst \left(k_BT \right)\frac{\partial}{\partial \alpha }\ e^{-\beta F}}{\int \mathcal{D} \*u \int \mathcal{D}
 f \int\mathcal{D} \mst e^{-\beta F}}\nonumber \\
 &= \frac{k_B T}{Z} \frac{\partial Z}{\partial \alpha}=k_B T\frac{\partial \ln Z}{\partial \alpha}.
 \label{eq:deltaA}
\end{align}
We will evaluate this expression at $\alpha=0$ to understand our simulations with
tension-free periodic boundary conditions.

We now simplify $Z$ by integrating out in-plane phonons so that our result
can be expressed in terms of $\mst(\*x)$ and the flexural phonon field $f(\*x)$. We Fourier
transform the free energy of Eq. 3, using the conventions
$\xi(\mathbf{x})=\sum_{\mathbf{q}} \xi(\mathbf{q}) e^{i \mathbf{q}
  \cdot \mathbf{x}}$
and
$\xi(\mathbf{q})=\frac{1}{A_0} \int d^{2} x \xi(\mathbf{x}) e^{-i
  \mathbf{q} \cdot \mathbf{x}}$, and then separate the (nonlinear) strain tensor into $\*q=0$ and $\*q \neq 0$
components.
\begin{equation}
u_{ij}(\*x)= u_{ij}^0 + A_{ij}^0 +\sum_{\*q\neq 0}\left( \frac{i}{2}  \left(q_i u_j(\*q) + q_j u_i(\*q)\right) +A_{ij}(\*q)\right)e^{i \*q \cdot \*x},
\end{equation}
where
$A_{ij}(\*x)= \frac{1}{2} \left( \frac{\partial f}{\partial x_i}
  \frac{\partial f}{\partial x_j} \right)$.
Since the Fourier transform of the pressure term only depends on
$u_{ij}^0$, we only need to consider the $\*q=0$ modes of the terms in
the free energy that depend on $u_{ij}$ to evaluate
Eq. \ref{eq:deltaA} as a function of $f$ and $\mst$. Note that $u_{ij}^0$ has three independent degrees of freedom, unlike $\frac{1}{2} \left( \partial_i u_j + \partial_j u_i \right)$ evaluated at finite wavevector.

Upon defining
\begin{equation}
\Psi(\*x)= \mst(\*x)^2,
\end{equation}
the $\*q=0$ mode of the free energy (suppressing the pure Ising terms, which do not depend on $u_{ij}$) is 
\begin{equation}
F^0= A_0 \left(\mu \left(u_{ij}^0+A_{ij}^0\right)^2 + \frac{\lambda}{2} \left(u_{kk}^0+A_{kk}^0 \right)^2+ g \Psi(0) \left(u_{kk}^0 +A_{kk}^0 \right)- \alpha u_{kk}^0 \right). 
\end{equation}

We now shift $u_{ij}^0$, defining
$u_{ij}^0= w_{ij}^0-A_{ij}^0-\frac{g \Psi(0)- \alpha}{2
  (\mu+\lambda)}\delta_{ij}$ \footnote{The corresponding expression in ref. \cite{plummer-PRE-102-033002-2020}, Eq. 24, is missing a factor of 1/2, and should instead read $\bar{u}_{\alpha \beta}^0=-A_{\alpha \beta}^0 + \frac{\Omega_0 c(0)}{2} \delta_{\alpha \beta}$. This change only produces corrections of order $\Omega_0^2$, which are dropped in this paper.}.
This change of variables produces a quadratic function of $w_{ij}^0$ that is easily integrated out. Our
remaining $\*q=0$ free energy, following the integration over in-plane
phonons, is
\begin{equation}
F^0= \frac{A_0}{2 (\mu+\lambda)} \left( - g^2 \Psi(0)^2 + 2 \alpha g \Psi(0) - \alpha^2\right)+ \alpha A_0 A_{kk}^0.
\end{equation}
Equation \ref{eq:deltaA} evaluated at $\alpha=0$ then gives
\begin{equation}
\langle \delta A \rangle = - \frac{g A_0 \langle \mst^2 \rangle}{\mu+\lambda}- \frac{A_0}{2} \left\langle \frac{\partial f}{\partial x_k} \frac{\partial f}{\partial x_k}\right \rangle,
\end{equation}
where $A_0$ is the projected area of the buckled system at $T=0$.

A complete understanding of the behavior of $\langle f(\*q) f(-\*q) \rangle$ and $\langle \mst(\*q) \mst(-\*q) \rangle$ would require evaluating the thermal averages with the effective free energy given by Eq. 4. However, as discussed in the main text, this model is still able to shed light on the simulation data for $\langle A\rangle/A_0$ in Fig. 4 and the observed divergence of the coefficients of thermal expansion. 

%\begin{align}
%\alpha_T &= \frac{1}{A_0} \frac{dA}{dT}= - \frac{d}{dT} \left(\frac{g\langle \mst^2 \rangle}{\mu+\lambda}\right)- \frac{d}{dT}\left\langle \frac{1}{2} \left(\frac{\partial f}{\partial x_i}\right)^2\right \rangle .
%\end{align}
%If we assume only a weak temperature dependence in the microscopic elastic parameters, $g$ and $\mu +\lambda$, we have
%\begin{align}
%\alpha_T=  -\frac{g}{\mu+\lambda} \frac{d \langle \mst^2 \rangle }{dT}- \frac{d}{dT}\left\langle \frac{1}{2} \left(\frac{\partial f}{\partial x_i}\right)^2\right \rangle .
%\end{align}
%The second term has the same form as the coefficient for thermal expansion of a pristine membrane, given by Eq. \AP{2} of the main text. 

\section{Stiffening of puckered surfaces for $T=0$}\label{andrej}
Using the techniques of ref. \cite{kosmrlj-PRE-88-012136-2013}, we argue here that the corrugations associated with antiferromagnetically buckled impurities lead to an increase in the effective membrane bending
rigidity at $T=0$. Following shallow shell theory \cite{koiter2009wt}, we assume
that there is a zero energy reference state described by
\begin{equation}
\*{r}_0(x_1,x_2)=\left(x_1, x_2, h(x_1, x_2) \right).
\end{equation}
In refs. \cite{kosmrljsphere, paulose}, among others, $h(x_1,x_2)$
describes a shallow section of a sphere, for example. For that case,
$h(x_1, x_2)=R \left(1 - \sqrt{1 - \frac{x_1^2}{R^2} -
    \frac{x_2^2}{R^2} } \right)$.
We instead assume that our reference metric corresponds to a
checkerboard pattern of up and down puckers, like an egg carton. Thus, deviations from this checkerboard egg carton ground state \cite{plummer-PRE-102-033002-2020} cost
energy. This description should be
accurate provided fluctuations are not able to invert any of the
buckled impurities, as will be the case for $T\ll T_c$.

To describe deviations from the reference state, which cost energy, we
separate the deviations into two components tangential to the
reference surface and one normal to the reference surface,
\begin{equation}
\*r=\*{r}_0+u_i \hat{\*{t}}_i^0+ f \hat{\*n}^0.
\end{equation} 

To extract the bending rigidity at $T=0$, we also allow for a position-dependent force/area difference $p(\*x)$ to act across the
membrane, with an energetic cost proportional to $f(\*x)$, the displacement in
the direction normal to the reference state. Ref. \cite{kosmrlj-PRE-88-012136-2013}
derives a self-consistent equation for the linear response of the
membrane in the presence of small $p$,
\begin{equation}
f(\*q) = \frac{p(\*q)}{\kappa q^4} - \frac{Y}{\kappa q^4} \sum_{\*q^\prime, \*q^{\prime \prime} \neq 0} \frac{  (\*q \times \*q^\prime)^2(\*q^\prime \times \*q^{\prime \prime})^2}{q^{\prime 4}}h(\*q-\*q^\prime)h(\*q^\prime-\*q^{\prime \prime}) f(\*q^{\prime \prime}).
\label{eq:selfconsistent}
\end{equation}
Following \cite{kosmrlj-PRE-88-012136-2013}, we assume that the term proportional
to $Y/\kappa$ can be treated as a perturbation, and solve iteratively by inserting
$f(\*q) = \frac{p(\*q)}{\kappa q^4} $ in the right hand side to get
the first order correction, linear in $p(\*q)$, 
\begin{equation}
f(\*q) \approx \frac{p(\*q)}{\kappa q^4} - \frac{Y}{\kappa q^4} \sum_{\*q^\prime, \*q^{\prime \prime} \neq 0} \frac{  (\*q \times \*q^\prime)^2(\*q^\prime \times \*q^{\prime \prime})^2}{q^{\prime 4}}h(\*q-\*q^\prime)h(\*q^\prime-\*q^{\prime \prime}) \frac{p(\*q^{\prime \prime})}{\kappa q^{\prime \prime 4}}.
\label{eq:firstcorr}
\end{equation}

The substitutions
\begin{align}
\*q^\prime&= \*q -\*q_1, \\
\*q^{\prime \prime}&= \*q- \*q_1- \*q_2,
\end{align}
lead to
\begin{equation}
f(\*q) \approx \frac{p(\*q)}{\kappa q^4} - \frac{Y}{\kappa^2 q^4} \sum_{\*q_1, \*q_2} \frac{  (\*q \times \*q_1)^2((\*q-\*q_1) \times \*q_2)^2}{(\*q-\*q_1)^4}h(\*q_1)h(\*q_2) \frac{p(\*q-\*q_1-\*q_2)}{(\*q-\*q_1-\*q_2)^4}.
\label{eq:firstcorrsimp}
\end{equation}
We now parametrize the deflections induced by the buckled impurities. The nonzero Fourier
modes compatible with the checkerboard reference state can be found by
direct calculation as in ref. \cite{plummer-PRE-102-033002-2020}, and are
\begin{equation}
\*B(m_1, m_2)=\left( \frac{(2 m_1+1) \pi}{n a_0} \hat{\*x} +  \frac{(2 m_2+1) \pi}{n a_0} \hat{\*y} \right),
\label{eq:reciprocal}
\end{equation}
where $m_1$ and $m_2$ are integers and $n a_0$ is the real space distance between impurities when $\epsilon=0$. 

Consider a long wavelength pressure
$p(\*x)=\sum_{\*q} p(\*q)e^{i \*q \cdot \*x}$, whose largest magnitude
wave vector component is much smaller than the smallest wave vector describing the checkerboard pattern. For this case, the correction term will
not contribute unless $\*q_1=-\*q_2$. With this condition,
Eq. \ref{eq:firstcorrsimp} simplifies to
\begin{equation}
f(\*q) \approx \frac{p(\*q)}{\kappa q^4}\left(1 - \frac{Y}{\kappa q^4} \sum_i \frac{  (\*q \times \*q_i)^4}{(\*q-\*q_i)^4}|h(\*q_i)|^2 \right) \approx \frac{p(\*q)}{\kappa q^4}\left( \frac{1}{1+ \frac{Y}{\kappa q^4} \sum_i \frac{  (\*q \times \*q_i)^4}{(\*q-\*q_i)^4}|h(\*q_i)|^2 }\right),
\end{equation}
where the sums are over wave vectors given by
Eq. \ref{eq:reciprocal}. We can now extract an effective bending
rigidity with a correction that is strictly positive for a given pressure wave vector $\*q$,
\begin{equation}
\kappa_R(\*q)  \approx \kappa \left( 1+\frac{Y}{\kappa q^4} \sum_i \frac{  (\*q \times \*q_i)^4}{(\*q-\*q_i)^4}|h(\*q_i)|^2 \right).
\end{equation}
Thus, at long wavelengths, the buckled defect texture stiffens
the membrane. The amount of stiffening depends on the direction and
wavelength of the pressure being applied, as well as the details of
the buckled defect structure encoded in $h(\*q)$. We can simplify the
expression further by applying the assumption $\*q \ll \*q_i$ for all
$\*q_i$, which leads to
\begin{equation}
\kappa_R(\*q) \approx \kappa \left( 1+\frac{Y}{\kappa} \sum_i \sin^4\theta_i |h(\*q_i)|^2 \right).
\end{equation}
where $\theta_i$ is the angle between $\*q$ and $\*q_i$. 

If we assume that the buckled dilation height field is well approximated by the eigenvectors derived in Sec. \ref{negative} using the smallest four wave vectors given by Eq. \ref{eq:reciprocal} as a basis, 
\begin{equation}
h(x,y)=h_0 \cos\left( \frac{\pi x}{n a_0} \right) \cos\left( \frac{\pi y}{n a_0} \right),
\end{equation}
for positive dilations, and
\begin{equation}
h(x,y)=h_0 \sin\left( \frac{\pi x}{n a_0} \right) \sin\left( \frac{\pi y}{n a_0} \right),
\end{equation}
for negative dilations, we can estimate the size of the correction. Upon carrying out an angular average over possible orientations of $\*q$, we have
\begin{equation}
\kappa_R \approx \kappa \left( 1+\frac{Y h_0^2}{16 \kappa} \langle \sum_i \sin^4\theta_i \rangle \right) \approx \kappa \left( 1+\frac{3 Y h_0^2}{32 \kappa} \right). 
\end{equation}
Close to the inextensible limit, we have $h_0^2 \sim \Omega_0$, the change in the surface area due to the dilation. If we are instead close to the buckling transition, $h_0^2 \sim a_0^2 (\gamma-\gamma_c$) \cite{plummer-PRE-102-033002-2020}. For the parameters used in our simulations, $Y=400/3, \kappa=1,$ and $h_0$ is measured to be approximately $0.4 a_0$. In this case,
\begin{equation}
\frac{3 Y h_0^2}{32 \kappa} \approx 12.5 \frac{h_0^2}{a_0^2} \approx 2.
\end{equation}
Therefore, according to this argument, we expect significant increases to the bending rigidity of the membrane due to the corrugated background of buckled dilations. Since this correction is large, violating the assumptions of our perturbation theory, we would need a more sophisticated calculation to estimate its magnitude more accurately. 

\section{Buckling threshold and eigenvectors for negative dilations}\label{negative}
The
continuum theory introduced in ref. \cite{plummer-PRE-102-033002-2020} to
treat positive dilations at $T=0$ can also be applied to
the case of negative dilations. We present these results here for
checkerboard arrays composed of either positive or negative
dilations. A high level overview of the calculation is given below; detailed explanations of the steps can be found in
ref. \cite{plummer-PRE-102-033002-2020}.

We start with the energy in the continuum limit minimized with respect to in-plane displacements,
\begin{equation}
E=\frac{1}{2}\int^\prime d^2r \bigg(\kappa \left(\nabla^2 f\right)^2 + Y\left(\frac{1}{2} P_{\alpha \beta}^T \partial_\alpha f \partial_\beta f\right) ^2 -Y\frac{\Omega_0}{2}P_{\alpha \beta}^T \partial_\alpha f \partial_\beta f c(\*r)\bigg),
\end{equation}
where the density of dilational impurities at positions \{$\*r_j$\} is given by $c(\*r)= \sum_j \delta^2(\*r-\*r_j)$ \cite{plummer-PRE-102-033002-2020}. Note that the area added/subtracted by each dilation is explicitly included in the elastic model here, in contrast to the phenomenological model of Sec. \ref{model}. We Fourier transform the height field, noting that the displacements are real ($f^*(\*q)=f(-\*q)$). We can now consider a particular pattern of defect buckling
(ferromagnetic, checkerboard, striped, etc.). For each pattern, there
is a subspace of wave vectors compatible with its periodicity that can
be used as a basis for the height field. For example, the height field
of the ferromagnetic state (equivalently, a single defect with
periodic boundaries) can be written as a sum over the reciprocal
lattice vectors of the defect superlattice.

As a first approximation, we truncate the wave vector basis so that it includes the vectors with the smallest magnitudes that allow the impurities to
couple to the flexural phonons. Since deformations are quite long range close to the buckling threshold \cite{plummer-PRE-102-033002-2020}, these longer
wavelength modes are particularly important.
%we are writing our height field in the momentum basis. Just like QM. 
We then write the quadratic contribution to the energy in matrix form,
and diagonalize the energy matrix to find the eigenvalues and
eigenvectors in reciprocal space. Next, we determine which eigenvalue first attains a negative value as the magnitude of $Y \Omega_0$
is increased-- this corresponds to the deformation for which the flat
state of the system is first unstable. The positive quartic term
ensures the existence of stable states with $f \neq 0$ beyond the
buckling threshold.

\begin{figure}[htp]
\begin{center}
\includegraphics[width=0.8\textwidth]{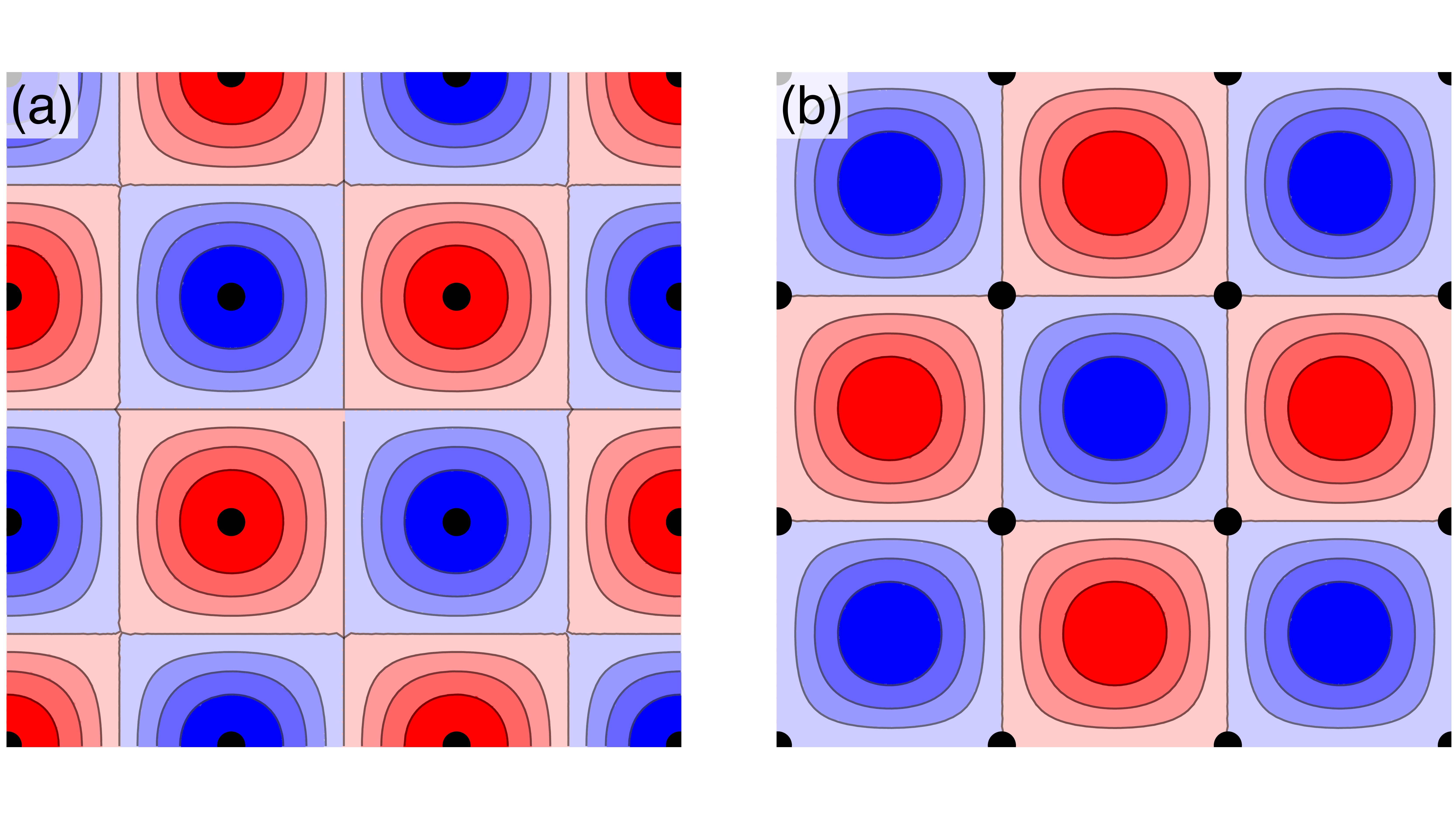}
\end{center}
\caption{{\label{fig:eigenvectors} Contour plots of the first unstable
    eigenvector in the continuum elastic analysis in real space using a basis of
    four wave vectors. Dilation nodes are located at the black dots. Negative deflections are blue, and positive deflections are red. The magnitude is arbitrary. (a) Positive $\Omega_0$. (b) Negative
    $\Omega_0$. }}
\end{figure}

We find the first eigenvectors to become unstable for both positive and
negative dilations in a checkerboard array in
Fig. \ref{fig:eigenvectors}. These eigenvectors are shifted by a uniform translation relative
to each other in position space, similar to what is seen in
simulations of $(0,2)$ arrays.

In this approximation, both eigenvectors become unstable at the same magnitude of
$\gamma_c$,
\begin{equation}
\gamma_c= \frac{Y|\Omega_0^c|}{\kappa}= 4 \pi^2
\end{equation}

\begin{figure}[htp]
\begin{center}
\includegraphics[width=0.8\textwidth]{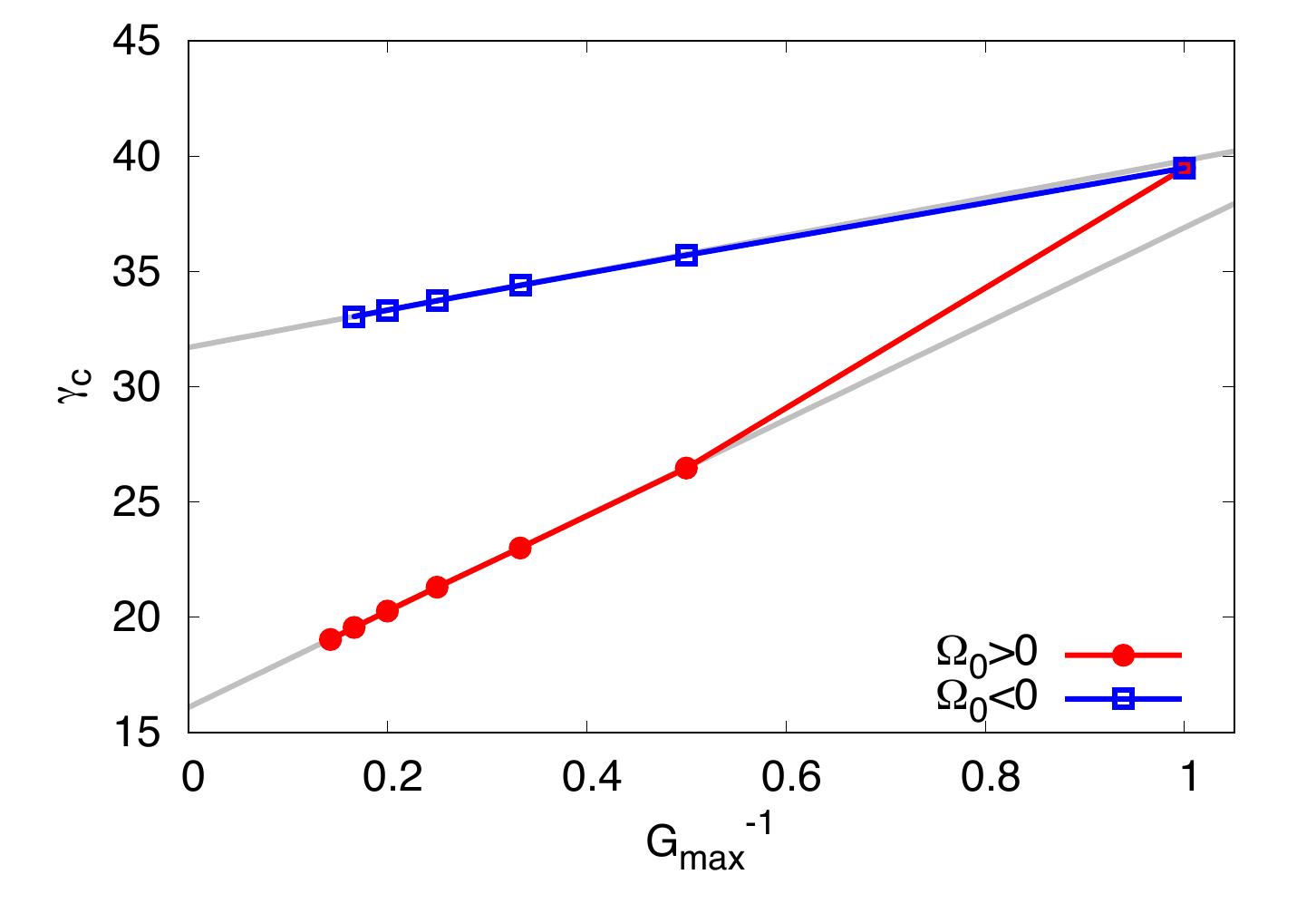}
\end{center}
\caption{{\label{fig:threshold} Variation of the buckling threshold
    $\gamma_c$ as we increase the number of Fourier modes included in
    the calculation. $G_\text{max}$ is the magnitude of the largest Fourier mode in the basis, measured in units of $2 \pi/d$, where $d$ is the separation between dilations in real space.  $G_{\text{max}}^{-1} \to 0$ corresponds to the continuum
    limit, in which case an infinite number of Fourier modes are included. Linear extrapolations to the continuum limit are shown by the grey lines.  }}
\end{figure}

To more accurately estimate the buckling thresholds, we increase the
number of Fourier modes included in the basis, and extrapolate to the
continuum limit. The results from these extrapolations are shown in
Fig. \ref{fig:threshold}:  Positive dilations buckle
earlier than negative dilations as the resolution of the calculation
is increased beyond the initial truncation, which agrees qualitatively with the simulations of $(0,2)$ superlattices presented in this paper. To pursue more quantitative agreement between the continuum theory and simulations
for negative dilations, we would need to find the buckling threshold
as a function of dilation separation, as we did for positive dilations
in Fig. 19 of ref. \cite{plummer-PRE-102-033002-2020}. We expect this continuum model is more accurate when dilations are further apart (and can be reasonably approximated as $\delta$-function strains).

\section{Molecular dynamics simulations}\label{sec:mdapp}
We study ``magnetic'' ordering of membranes with dilations that
provide positive and negative extra area ($\Omega_0>0$ and
$\Omega_0<0$ respectively) at finite temperatures with periodic
boundary conditions in $x$ and $y$ directions. The energy, adapted
from ref. \cite{seung-PRA-38-1005-1988} and described in the main text, is given by
\begin{equation}
  E=\frac{k}{2}\sum_{\langle i,j\rangle}(|\pmb{r}_{i}-\pmb{r}_j|-a_{ij})^2+\hat{\kappa}\sum_{\langle \alpha, \beta \rangle}(1-\pmb{n}_{\alpha}\cdot\pmb{n}_{\beta}),
\end{equation}
where $k$ is the spring constant, $\hat{\kappa}$ is the microscopic
bending rigidity, and $a_{ij}$ is the rest length between two
connected nodes, $i$ and $j$. The first sum is over connected nodes and the second
sum is over neighboring triangular planes.

For pristine membranes, the corresponding continuum Young's modulus is
$Y=4k/3$ and the continuum bending rigidity is
$\kappa=\hat{\kappa}$~\cite{plummer-PRE-102-033002-2020}. The
continuum size of the dilation is defined as $\Omega_0=4a_0^2\epsilon$
\cite{plummer-PRE-102-033002-2020}.  In
ref. \cite{plummer-PRE-102-033002-2020}, it was shown that
$\Omega_0= 4 a_0^2 \epsilon$ for positive dilations. In fact, this
relation holds for negative dilations as well, which can be confirmed
by observing that the argument of \cite{plummer-PRE-102-033002-2020}
does not depend on the sign of $\epsilon$, and with a numerical check.

We choose microscopic elastic parameters
$a_0=1, k=100, \hat{\kappa}=1, \epsilon=\pm 0.1$, corresponding to
dilation F\"oppl-von K\'arm\'an number
\cite{plummer-PRE-102-033002-2020}
$\gamma= \frac{Y |\Omega_0|}{\kappa}\approx 53$, and place neighboring
dilations two lattice spacing apart. For these parameters, positive
and negative dilation arrays are structurally similar.  The arrays are
dense and we are close enough to the buckling threshold
($\gamma_c \approx 21$ for positive dilations, $\gamma_c\approx 26$
for negative dilations) that the out-of-plane deformations of
neighboring dilations overlap strongly, producing strong interactions.

We run molecular
dynamics (MD) simulations with HOOMD~\cite{anderson2020hoomd} on
NVIDIA Tesla V100 GPU and we accumulated roughly 1 terabyte of
data. The simulations at finite temperatures are carried out within
the $NPT$ ensemble (fixed number of particles $N$, pressure $P$, and
temperature $T$) at zero pressure. The NPT integration is carried out
via Martyna-Tobias-Klein barostat-thermostat with a time step
$dt=0.001$.  For zero-temperature structural relaxation, we use Fast
Inertial Relaxation Engine (FIRE) algorithm~\cite{fire-algo,
  anderson2020hoomd} with a step size $dt=0.005$, force and energy
convergence criteria of $10^{-6}$ and $10^{-10}$, respectively.

We simulate membranes with $L_N\times L_N$ nodes on a square lattice
with $L_N=24, 36, 48, 60,$ $80, 96, 120, 160, 192$; these values
correspond to membranes with a total number of sites ranging from 576
to 36864. Dilations are placed on a square lattice two unit spacings
($2a_0$) apart along $x$ and $y$ directions to form a $(0,2)$
tiling. We denote $L_{\rm I}\equiv L_N/2$ as the number of ``spins''
(buckled sites) along one axis which is the linear dimension of our
mechanical antiferromagnet.

The coarse-grained model with no
dilations~\cite{seung-PRA-38-1005-1988} has been used to study the
mechanical and thermal properties of 2D materials (e.g.
graphene)~\cite{zhang-JMPS-67-2-2014, bowick-PRB-95-104109-2017,
  yllanes-NatCom-8-1-2017, hanakata-EML-44-101270-2020,
  morshedifard-JMPS-149-104296-2021}. We fix $\hat{\kappa}=1.0$ and
$k=100$ and vary $T$ from 0.1 to 0.4. At each $T$, snapshots of
positions are taken every 10,000 steps and a total run of $10^7$ steps
is performed. For systems with $L_N>96$ we perform $2\times10^7$
steps.  For each $T$ we perform at least 10 independent runs---near
$T_c$, we perform 15 to 50 independent runs. Longer runs and more
independent runs are required for larger systems. The first half of
the snapshots are discarded to allow for thermal equilibration.

\section{Equilibration}\label{sec:equilapp}
\subsection{Ground states at $T=0$}
To probe the ground state of our mechanical analog of an Ising AFM, we
first performed structural relaxation on a system with $L_N=4$ which
has 2x2 dilations, with energy minimization and cell optimization at
$T=0$ via the Fast Inertial Relaxation Engine (FIRE) algorithm. The
cell optimization allows the lattice parameters to deform during
minimization to achieve a zero stress state. We initialize the buckled
sites of positive dilations ($\Omega_0>0$) with two different initial
conditions (i) AFM and (ii) FM with some noise. In both cases, the
final relaxed configurations indicate antiferromagnetic buckled order,
as found by ref. \cite{plummer-PRE-102-033002-2020}.

For $\Omega_0<0$, we also initialize the buckled background sites with
AFM and FM states.  We find that the relaxed state for the AFM
initialization has a total energy per ``spin'' $E/N_I=0.6475$. In
contrast, when we initialize the system with an FM state, the final
configuration has 3 ``spins'' pointing in the same direction, and
$E/N_I=0.8575$. Thus, for both positive and negative dilations the AFM
state has the lower energy. In the following section, we also find a
robust AFM ground state by quenching a large system ($L_N=96$, 48x48
dilations) initialized with a disordered configuration to a low
temperature ($T=0.1$) using molecular dynamics.
\subsection{Thermal equilibration for $T>0$}
\label{sec:equil}
\begin{figure}[htp]
\begin{center}
\includegraphics[width=0.8\textwidth]{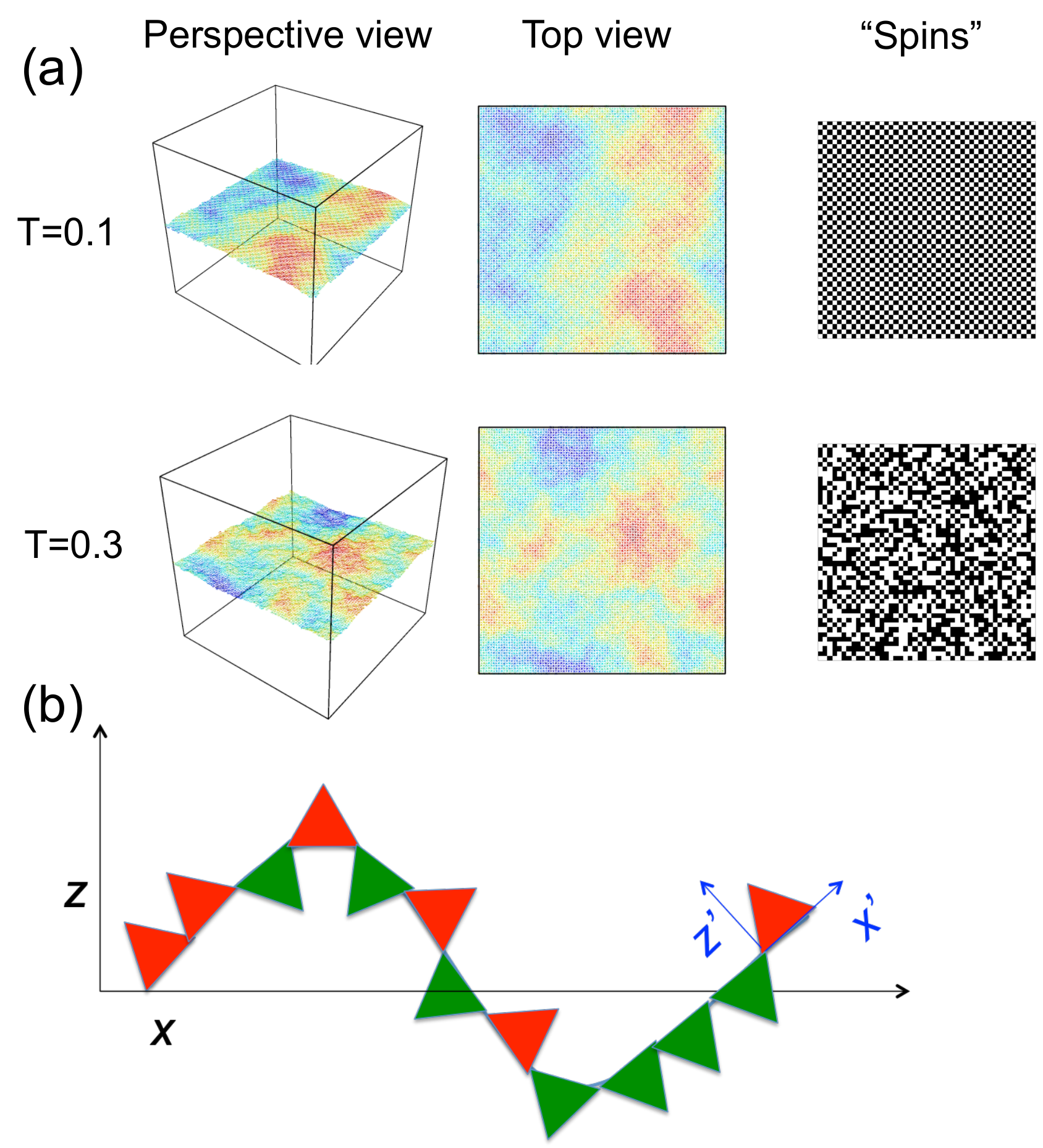}
\end{center}
\caption{(a) Snapshots of all nodes at low and high temperatures and
  the corresponding ``spin'' configurations. The heat map shows the
  height of each node relative to the $xy$-plane and scaled by the
  maximum and the minimum value at a given snapshot. In the spin
  configuration snapshots, black pixels represent upward buckles and
  white pixels represent the downward buckles (b) Schematic side view
  of a chain of buckled sites (spins). Because of the curvatures due
  to thermal fluctuations, a local frame formed by neighboring nodes
  is used to define up and down spins (buckling direction). }
\label{fig:spinConfigurations}
\end{figure}
In this section, we investigate the equilibration of puckered
membranes at low and high temperatues, $T=0.1, 0.3$. To test
equilibration, we compare systems initialized with two different
configurations: (i) the known AFM ground state at $T=0$ and (ii) a
disordered state. Here, we use 48x48 dilations hosted on a 96x96
square lattice ($L_N=96$). For initialization with the AFM
configuration, we first set the heights of impurities following AFM
ordering and then add some small random deformations to all nodes. We
then performed energy and stress minimizations at $T=0$ using the FIRE
algorithm. MD is implemented following this relaxation.  For
disordered initializations, we set the heights of all nodes with
random numbers.

Figure~\ref{fig:spinConfigurations}(a) shows snapshots of positions of
\emph{all} nodes at $T=0.1$ and $T=0.3$. The heat map indicates the
position of a node relative to the reference $xy$-plane ($z=0$). Based
on the height profiles, we see that thermal fluctuations excite the
out-of-plane mountains and valleys associated with flexural
deformations, also present in membranes without dilations. Since the
membrane is no longer perfectly flat at non-zero $T$, we need to use a
local reference frame (formed by local neighbors) to identify whether
a dilation is in fact buckled up or down (see
Fig.~\ref{fig:spinConfigurations}(b)). To assign up and down spins to
buckled sites at finite temperature, we use the nodes' positions
relative to the local planes formed by their neighbors. This procedure
accounts for thermal fluctuations relative to the zero plane, which
can be much larger than the typical dilation buckling amplitude in
this highly compressible medium (approximately $0.4 a_0$).

% We assign a positive spin
%variable ($s=+1$) to a site with a position above the the plane formed
%by its neighbors and a negative spin ($s=-1$) otherwise.

Using the filtering described above, we are able to map the positions
of the buckled sites to spin configurations (see
Fig.~\ref{fig:spinConfigurations}(a)). To study magnetic ordering in
this system, we use the standard stagger operator to calculate the
staggered magnetization per spin
\begin{equation}
m_{\rm st}=\frac{1}{N_{\rm I}}\sum_i^{N_{\rm I}}s_i(-1)^{x_i+y_i}
\end{equation}
where $N_{\rm I}$ is the total number of dilations, $s_i$ is the
dilation's ``spin'' and $x_i, y_i$ are the site indices on a 2D square
lattice.

\begin{figure}[htp]
\begin{center}
\includegraphics[width=0.8\textwidth]{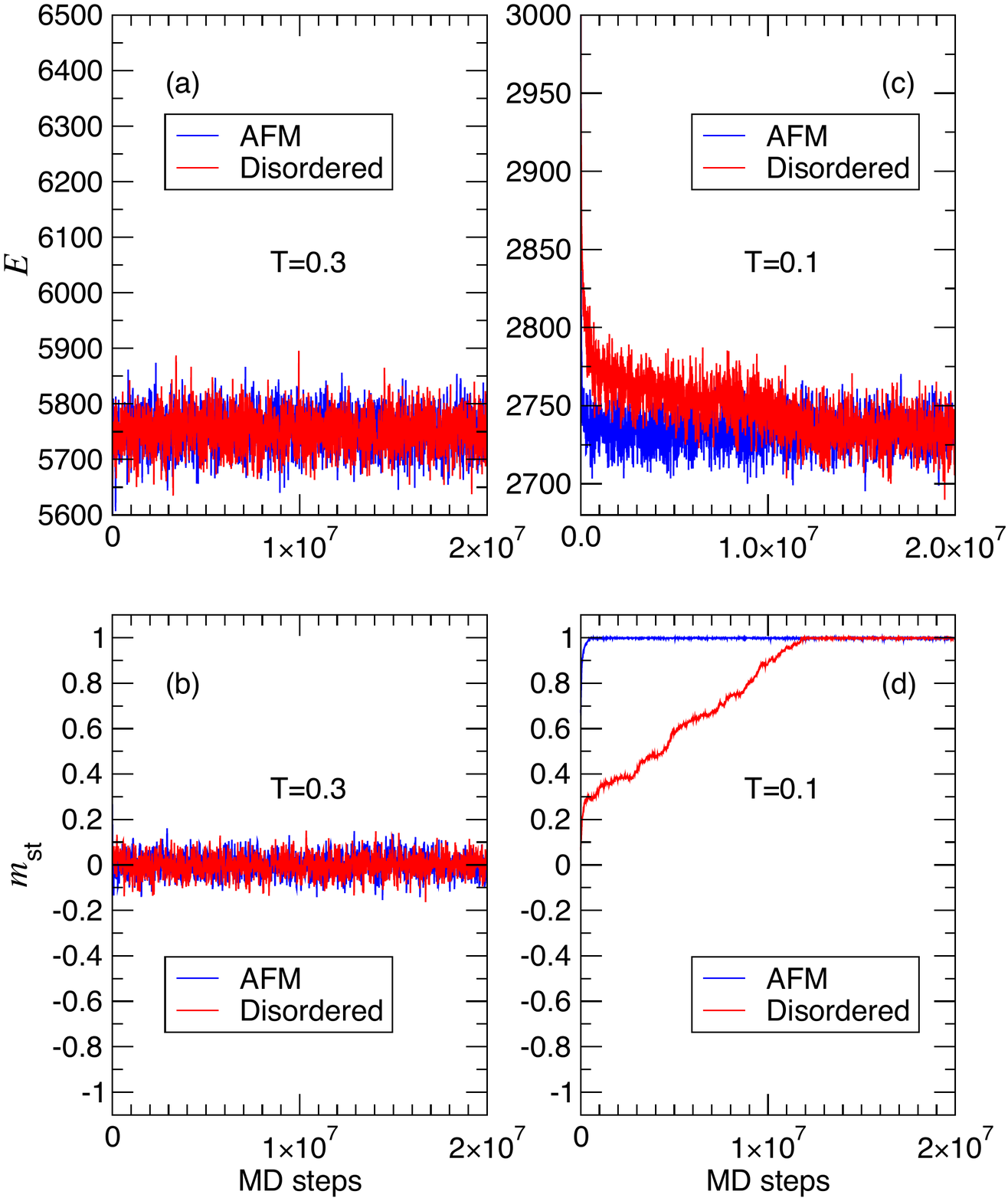}
\end{center}
\caption{ The total potential energy $E$ and the staggered
  magnetization per spin $\mst$ for membranes with $\Omega_0>0$ as a
  function of MD steps at high (a, b) and low (c, d) temperatures. The
  simulations were initialized in both a disordered state and an AFM
  configuration.}
\label{fig:equilibration_puckers}
\end{figure}

\begin{figure}[htp]
\begin{center}
\includegraphics[width=0.8\textwidth]{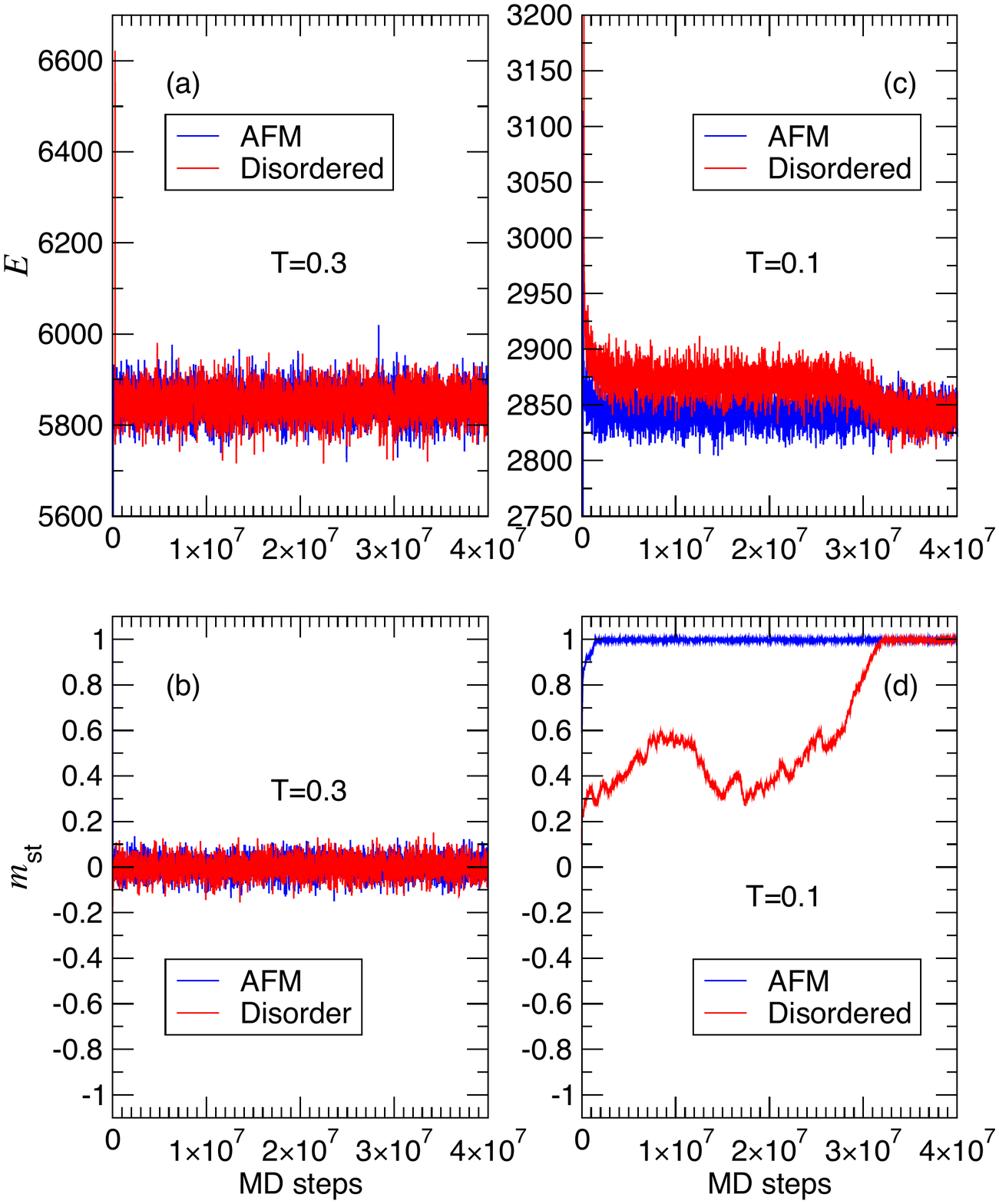}
\end{center}
\caption{ The total potential energy $E$ and the staggered
  magnetization per spin $\mst$ for membranes with $\Omega_0<0$ as a
  function of MD steps at high (a, b) and low (c, d) temperatures. The
  simulations were initialized in both a disordered state and an AFM
  configuration.}
\label{fig:equilibration_stitches}
\end{figure}

Next, we monitor the total potential energy $E$ and the staggered
magnetization per spin $m_{\rm st}$ to check if the system is in
equilibrium. Figure~\ref{fig:equilibration_puckers}((a)-(d)) shows $E$
and $\mst$ as a function of time for systems prepared in both an AFM and
a disordered state.  At high $T$, we see $E$ and $\mst$ of the
disordered and AFM states equilibrate after $\sim5\times10^5$ MD
steps. At low $T$, the systems prepared with an AFM state relaxes
after $\sim5\times10^5$ MD steps, whereas the system prepared with a
disordered state relaxes after $\sim1\times10^7$ MD steps. Similar
results are found for membranes with negative dilations, shown in
Fig.~\ref{fig:equilibration_stitches}. We can see that for both cases
the AFM state is robust even at finite temperature, provided $T<T_c\simeq0.2$.

We could start with disordered states (high $T$) and quench the
systems to low $T$ to initialize all simulations; however, this would
take too many computing resources as the equilibration time is quite
long at low $T$.  To run $2\times10^7$ MD steps, it takes 1 hour using
one GPU or $\sim30$ hours using one CPU. Since we know that the ground
state at $T=0$ and the equilibrated state at low $T$
(well below the critical temperature) is a mechanical analog of an
Ising AFM state, we will use the AFM configurations (with some small noise)
as the initial conditions for all simulations.

\section{Critical exponents}\label{critall}
In any system of finite dimensions, the correlation length $\xi$
cannot exceed the system size $L$ and divergent quantities, such as
staggered antiferromagnetic susceptibility $\chi$ and specific heat
$C$, will reach a maximum at a pseudocritical temperature
$T_c(L)$. Note that unlike the usual incompressible nearest neighbor
2D Ising model we do not know the true critical temperature
($T_c(L\to \infty)$) in the thermodynamic limit. We use a standard
finite-size scaling (FSS) approach to extract critical exponents,
similar to methods used to determine critical exponents from Monte
Carlo (MC) simulation studies of Ising models and related
models~\cite{binder-RepProgPhys-1997, sandvik-AIP-2010}. Since
$\xi\sim |T-T_c|^{-\nu}$ and $\xi_{T=T_c(L)}\simeq L$, we can rewrite
the temperature dependence of susceptibility, specific heat, and
magnetization at $T_c(L)$ in terms of system size,
\begin{align}
\chi&\sim L^{\gamma/\nu},\nonumber\\
C&\sim L^{\alpha/\nu}\nonumber,\\
\sqrt{m^2}&\sim L^{-\beta/\nu}.\nonumber
\end{align}
Using this finite size scaling approach with extensive simulations for
$L_N=24, 36, 48, 60, 80, 96, 120, 160$ and 192 we obtain
$\gamma/\nu=1.741\pm0.062$, $\alpha/\nu=0.068\pm0.018$,
$\beta/\nu=0.127\pm0.038$, and $\nu=0.99\pm0.09$ for $\Omega_0>0$ and
$\gamma/\nu=1.684\pm0.061$, $\alpha/\nu=0.074\pm0.016$,
$\beta/\nu=0.111\pm0.036$, and $\nu=1.05\pm0.25$ for $\Omega_0<0$.

The critical exponents of membranes with positive and negative dilation arrays and the exact
critical exponents of the 2D Ising model are tabulated in
Table~\ref{table:table1}.
\begin{table}
\small
\caption{Critical exponents $\alpha, \beta, \gamma$, and $\nu$ for membranes with positive and negative dilations. The uncertainties include errors from fitting and statistical errors from measurements. }
\centering
\begin{tabular*}{0.8\textwidth}{@{\extracolsep{\fill}}|c|c|c|c|c|}
\hline 
& $\alpha$ & $\beta$ & $\gamma$ & $\nu$ \\
%\\ \\ [0.5ex]
\hline 
$\Omega_0>0$  &   $0.067\pm0.018$&$0.126\pm0.040$&$1.723\pm0.162$&$0.99\pm0.09$ \\
$\Omega_0<0$ &   $0.078\pm0.025$&$0.117\pm0.047$&$1.768\pm0.425$&$1.05\pm0.25$ \\
2D Ising    & 0(log) &  1/8    & 7/4    & 1 \\
%\\ [1ex] 
\hline
\end{tabular*}
\label{table:table1}
\end{table}
%Puckers  &   $0.11\pm0.02$&$0.091\pm0.001$&$1.76\pm0.2$&$0.99\pm0.10$ \\
Note that we \emph{directly }measured $\gamma/\nu$, $\alpha/\nu$, and
$\beta/\nu$ using finite size scaling. In order to calculate
$\gamma, \alpha,$ and $\beta$, we need to use $\nu$ obtained from other
measurements. We use a power law fitting in the form of
$L=c_0|T_c(L)-T_c(\infty)|^{-\nu}$ to extract $\nu$. We excluded the two smallest systems ($L_N=24, 36$) in all
fittings. Details of the fitting for each critical exponent will be
discussed in the subsections below. %We use the same fitting procedures
%for membranes with negative dilations ($\Omega_0<0$).
\subsection{Susceptibility $\chi$ and critical exponent $\gamma$}\label{critgamma}
\begin{figure}[htp]
\begin{center}
\includegraphics[width=0.8\textwidth]{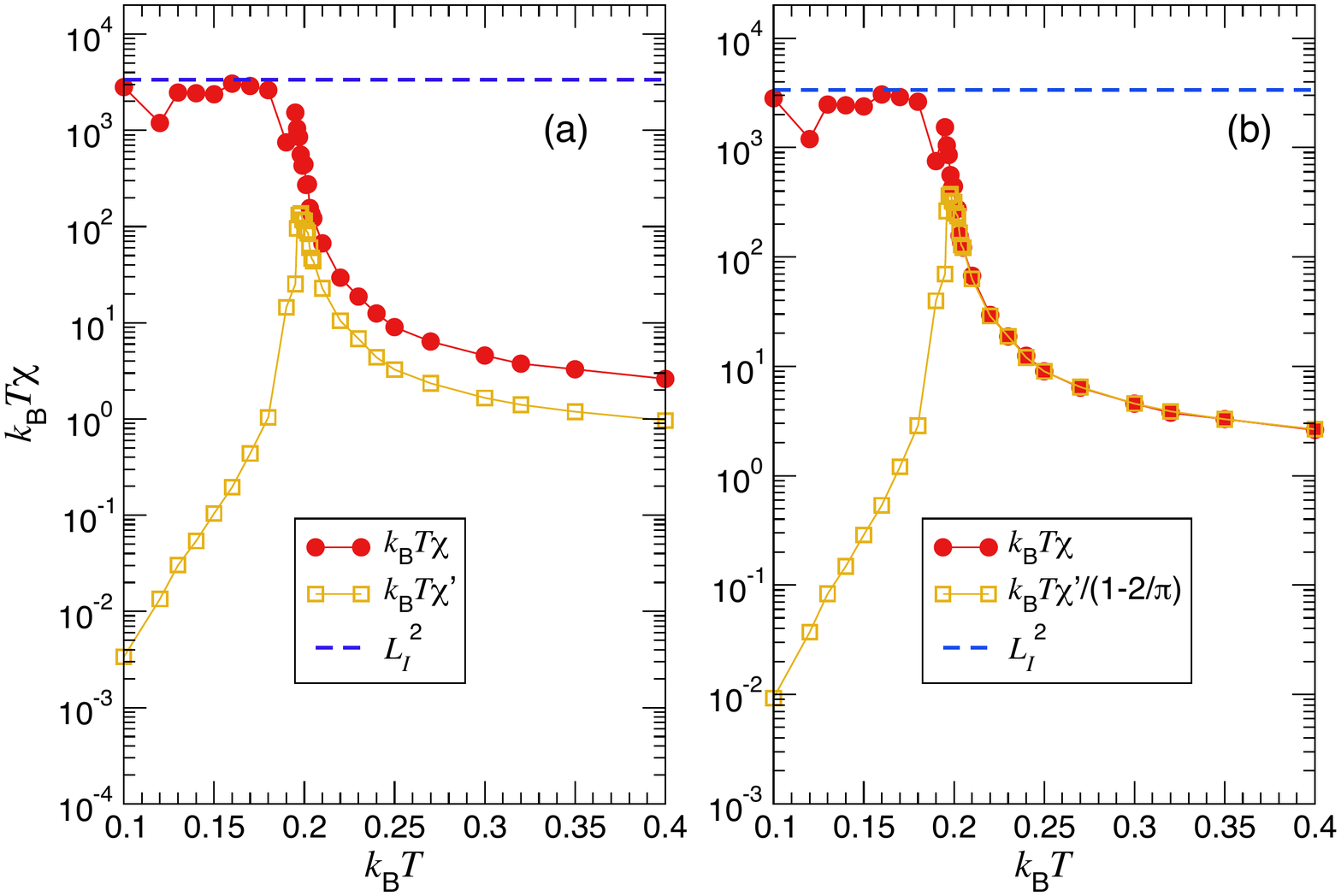}
\end{center}
\caption{(a) The normalized staggered susceptibility for positive
  dilations
  $k_{\rm B}T\chi=N_{\rm I}\left(\langle \mst^2\rangle -\langle
    \mst\rangle^2\right)$
  and the closely related quantity
  $k_{\rm B}T\chi'=N_{\rm I}\left(\langle \mst^2\rangle
    -\langle|\mst|\rangle^2\right)$
  as a function of temperature $T$ for a membrane of size
  $L_N=120$. The total number of up/down puckers
  $N_I=L_I\times L_I=60\times60$. We can see $k_{\rm B}T\chi$
  converges to the system size $N_I=L_I\times L_I$ at low $T$. (b) For
  $T>T_c$, we see that $k_{\rm B}T\chi'/(1-2/\pi)$ closely tracks
  $k_{\rm B}T\chi$~\cite{binder-RepProgPhys-1997}.  }
\label{fig:compareXi}
\end{figure}
%In studies of antiferromagnetic critical behavior, it is common to study the
%staggered susceptibility $\chi$ and the specific heat $C$. The quantities $\chi$
%and $C$ can be calculated by measuring fluctuations in ``spins'' (up/down buckling) and
%energy, respectively. We use the staggered magnetization per spin as
%the order parameter
%\begin{equation}
%  \mst=\frac{1}{N_I}\sum_is_i(-1)^{n_x+n_y},
%\end{equation}
%where $s_i=\pm1$ is the spin on site $i$ and $n_x, n_y$ are the site
%indices on a 2D square lattice. Up spin (+1) corresponds to a
%buckled site with a position above the plane formed by its
%neighbors.
 The staggered susceptibility is given by
\begin{equation}
\chi = \frac{N_{\rm I}}{k_{\rm B}T}\left(\langle \mst^2\rangle -\langle \mst\rangle^2\right),
\end{equation}
where $N_{\rm I}$ is number of dilations. True spontaneous symmetry breaking
can occur in the thermodynamic limit only. For a finite system, there
is a probability for the spontaneous staggered magnetization
$\langle \mst\rangle$ to flip after a long finite time (unless we
apply a small symmetry-breaking staggered field) and thus 
$\langle \mst\rangle$ is zero for all $T$.
%This can be seen by taking the double limits
%\begin{equation}
%m_{\rm sp}=\lim_{h\to 0}\langle \mst\rangle_{T, h}=0\quad \forall N_I.
%\end{equation}

Thus, for any finite system under zero external field 
$k_{\rm B}T\chi=N_I\langle\mst^2\rangle$. Consequently,
$k_{\rm B}T\chi$ is a monotonically increasing function that converges to $N_I$ as
$T\to 0$~\cite{binder-RepProgPhys-1997}. As suggested by
Binder~\cite{binder-RepProgPhys-1997}, we can use $|\mst|$ to
calculate a closely related quantity
\begin{equation}
\chi' = \frac{N_{\rm I}}{k_{\rm B}T}\left(\langle \mst^2\rangle -\langle |\mst|\rangle^2\right),
\label{eq:chi'}
\end{equation}
which is better behaved. Below $T_c$, $\chi'$ converges to the true
susceptibility in the thermodynamic limit $L_N\rightarrow \infty$. The
quantity $\chi'$ is different from the true susceptibility for $T>T_c$
by a simple multiplicative factor
$1-2/\pi$~\cite{binder-RepProgPhys-1997}.
\begin{figure}[htp]
\begin{center}
\includegraphics[width=0.8\textwidth]{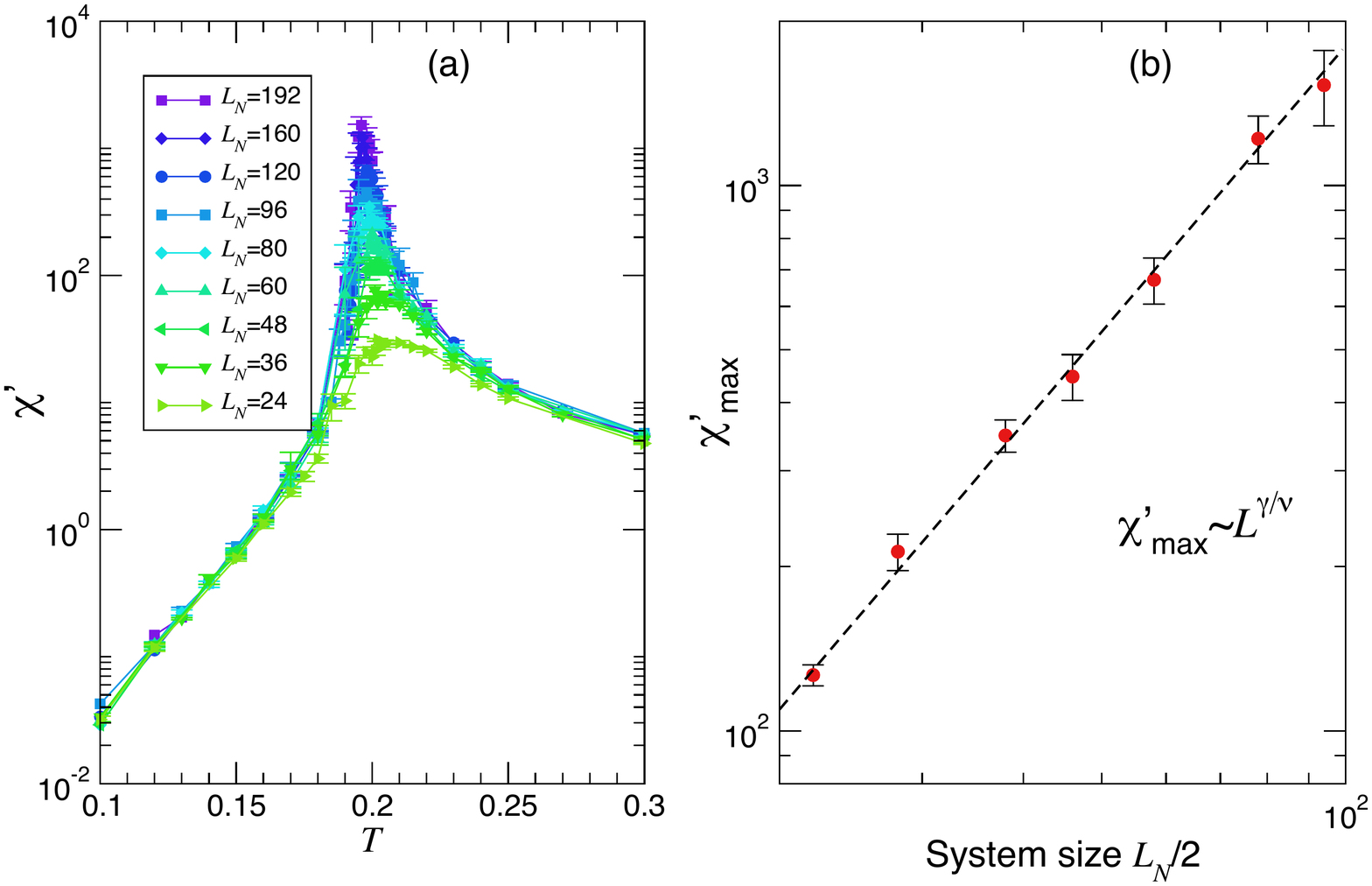}
\end{center}
\caption{The function $\chi'$ given by Eq.~\ref{eq:chi'} for membranes with $\Omega_0>0$ as a function
  of temperature $T$ for different linear system sizes $L_N$. $\chi'$
  increases with increasing system size and the location of pseudocritical temperature $T_c(L)$, where $\chi'$ is at its maximum, decreases with
  increasing system size. Note that for our (0, 2) dilation configuration
  on a square lattice, the linear system size of the spins is
  $L_I=L_N/2$. (b) Log-log plot of susceptibility peaks
  $\chi'_{\rm max}$ as a function of linear system size $L_N/2$. The
  data points are fitted with a power law function and we find
  $\gamma/\nu=1.741\pm0.062$.}
\label{fig:Xi-T_and_XiMax-L}
\end{figure}
\begin{figure}[htp]
\begin{center}
\includegraphics[width=0.8\textwidth]{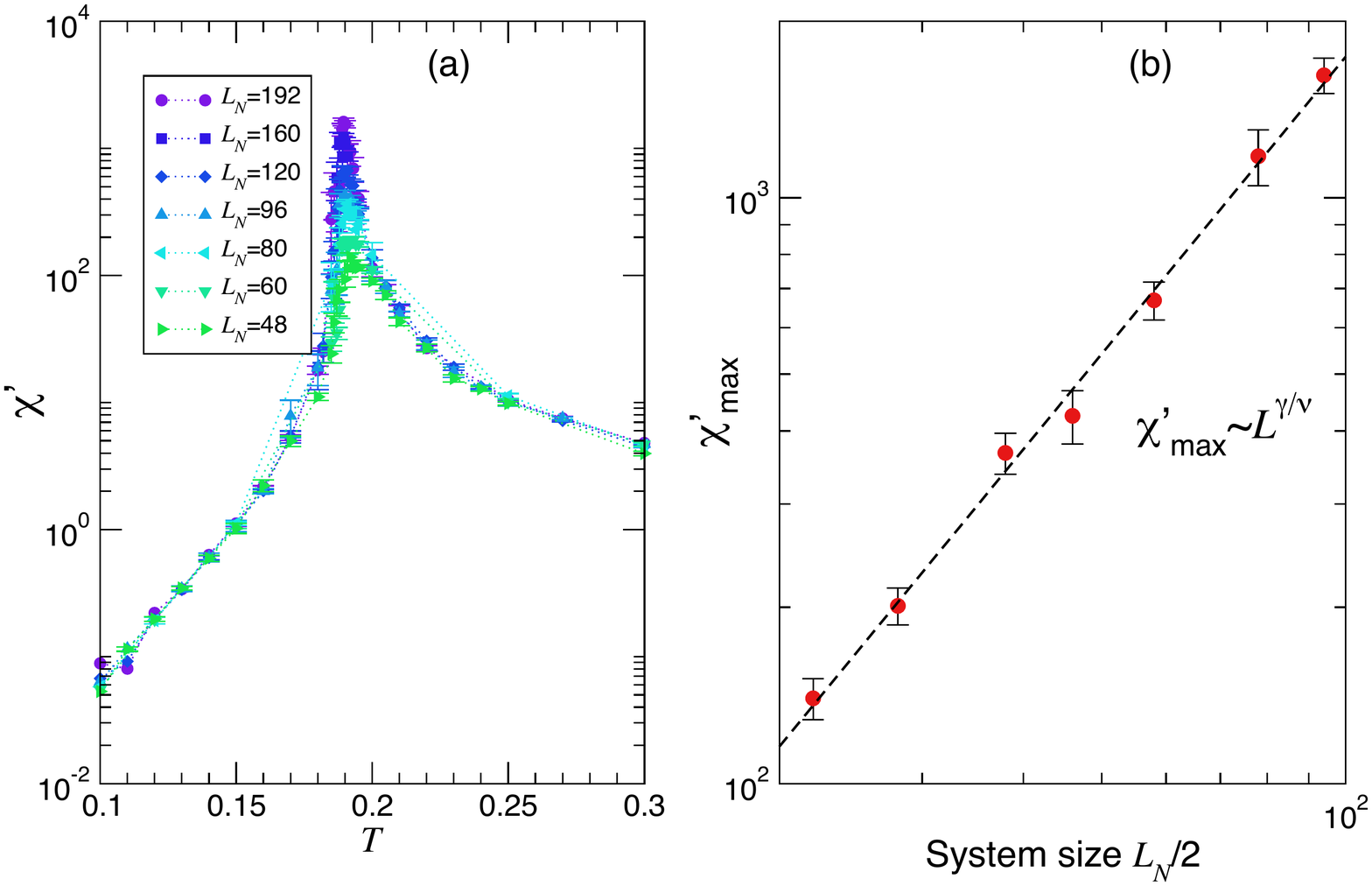}
\end{center}
\caption{The function $\chi'$ given by Eq.~\ref{eq:chi'} for membranes with $\Omega_0<0$
  as a function of temperature $T$
  for different linear system sizes $L_N$.
  (b) Log-log plot of susceptibility peaks $\chi'_{\rm
    max}$ as a function of linear system size
  $L_N/2$.
  The data points are fitted with a power law function and we find
  $\gamma/\nu=1.684\pm0.061$.}
\label{fig:Xi-T_and_XiMax-L_stitches}
\end{figure}
We compare $k_{\rm B}T\chi$ and $k_{\rm B}T\chi'$ obtained from MD
simulations of a membrane with $N_I=60\times60$ positive dilations in
Fig.~\ref{fig:compareXi}.  We indeed find that $k_{\rm B}T\chi$
converges to the system size at low temperatures ($T<T_c$) and
$k_{\rm B}T\chi'\simeq k_{\rm B}T\chi(1-2/\pi)$ at high
temperatures. Since $\chi'$ is only different by a simple
multiplicative factor, many Monte Carlo studies of spin systems use
$\chi'$ to extract the critical exponent
$\gamma$~\cite{binder-RepProgPhys-1997, sandvik-AIP-2010}.  We use $\chi'$ to extract susceptibility exponents for
membranes with $\Omega_0>0$ in Fig.~\ref{fig:Xi-T_and_XiMax-L}. We see that the peak
increases with increasing system size whereas the location of the peak
decreases with increasing system size. The maximum value of $\chi'$ as a
function of linear system size $L_I\equiv L_N/2$ in a log-log plot is
shown in Fig.~\ref{fig:Xi-T_and_XiMax-L}(b). By fitting the data to a
power law function we obtain $\gamma/\nu=1.741\pm0.062$.  Similar
trends are found for $\chi'$ in membranes with negative dilations, as
shown in Fig.~\ref{fig:Xi-T_and_XiMax-L_stitches}. For negative
dilations we find $\gamma/\nu=1.684\pm0.061$. Using $\nu$ obtained
from FSS (see below), we obtain $\gamma_{\Omega_0>0}=1.72\pm0.16$ and
$\gamma_{\Omega_o<0}=1.77\pm0.42$.% which are close to the 2D Ising model $\gamma=7/4$.

%gamma=1.237075, nu=0.629971
%1.963
\subsection{Specific heat $C$ and critical exponent $\alpha$}\label{critalpha}
\begin{figure}[htp]
\begin{center}
\includegraphics[width=0.8\textwidth]{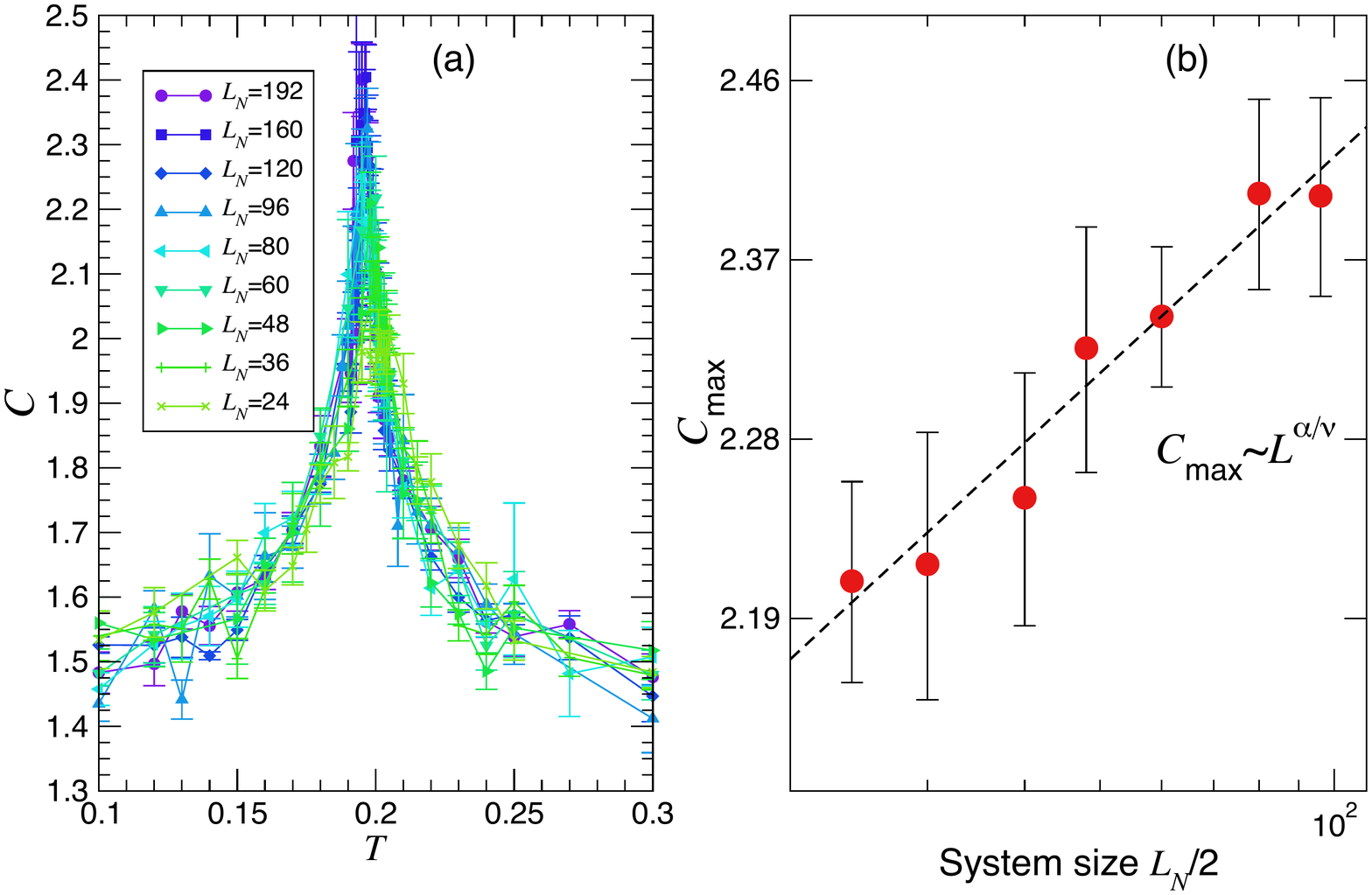}
\end{center}
\caption{Specific heat of membranes with positive dilations
  $\Omega_0>0$ as a function of $T$ for nine different linear system
  sizes $L_N$. (b) Log-log plot of peaks of specific heat
  $C_{\rm max}$ as a function of linear system size $L_N$. The data
  points are fitted with a power law function and we find
  $\alpha/\nu=0.068\pm0.018$.}
\label{fig:C-T_and_CMax-L}
\end{figure}
\begin{figure}[htp]
\begin{center}
\includegraphics[width=0.8\textwidth]{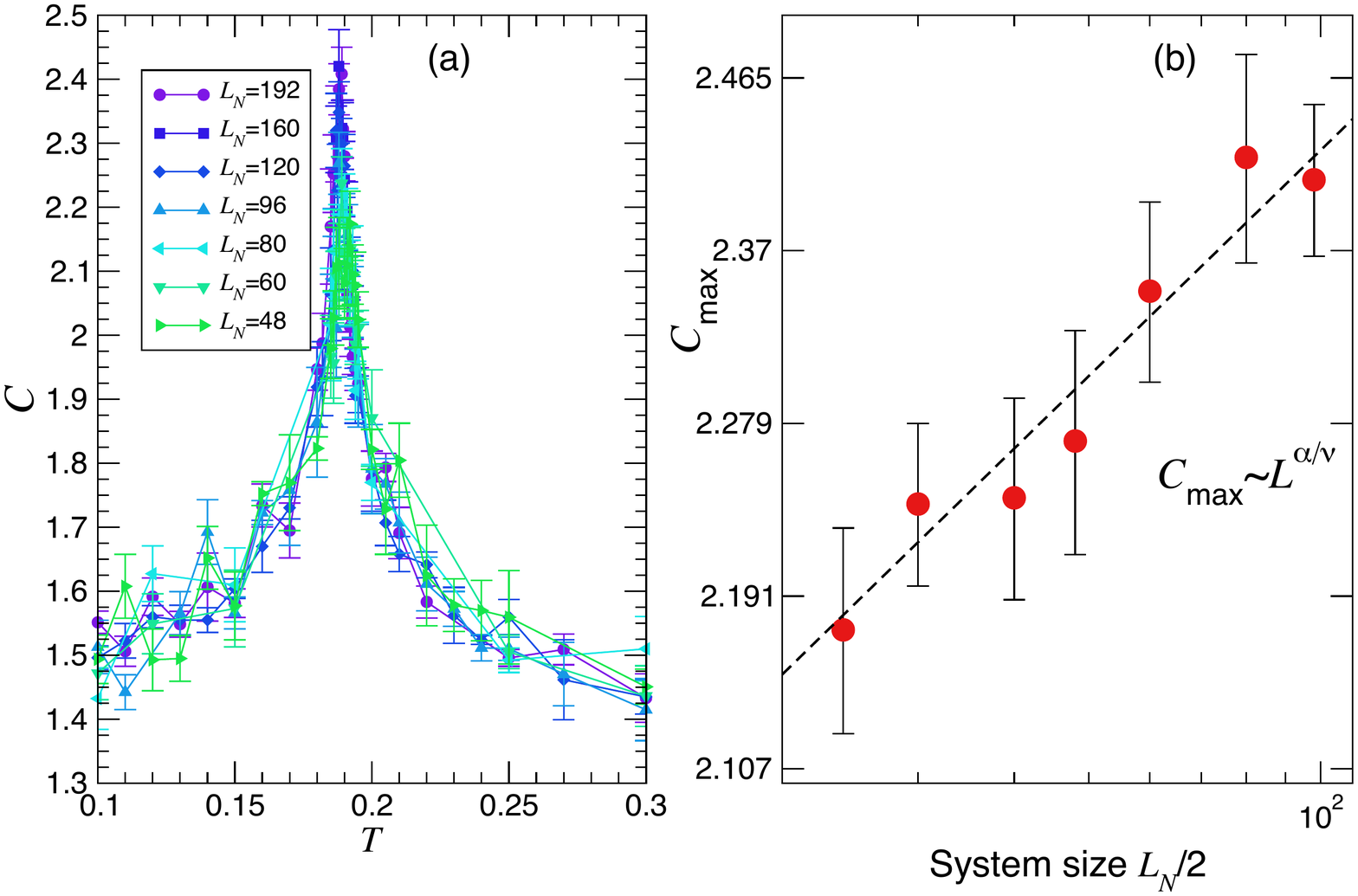}
\end{center}
\caption{Specific heat of membranes with $\Omega_0<0$ as a function of
  $T$ for different linear system sizes $L_N$. (b) Log-log
  plot of peaks of specific heat $C_{\rm max}$ as a function of linear
  system size $L_N$. The data points are fitted with a power law
  function and we find $\alpha/\nu=0.074\pm0.016$.}
\label{fig:C-T_and_CMax-Lneg}
\end{figure}
The average values of the energy and the square of the energy in
thermodynamic equilibrium are given by
\begin{align}
&\langle E\rangle = \frac{1}{Z}\sum_jE_je^{-E_j/k_{\rm B}T}\\
&\langle E^2\rangle = \frac{1}{Z}\sum_jE_j^2e^{-E_j/k_{\rm B}T},
\end{align}
where $Z$ is the partition function. Upon taking the derivative 
of $\langle E\rangle$ we find the specific heat for $N$ degrees of freedom:
\begin{equation}
NC\equiv\frac{\partial \langle E\rangle}{\partial T}=\frac{1}{k_{\rm B}T^2}(\langle E^2\rangle-\langle E\rangle^2), 
\end{equation}
where $C$ is the specific heat per node. We can calculate
$\langle E\rangle$ and $\langle E^2\rangle$ from MD simulations by
averaging over equilibrated configurations. 

Note in the simulations we measure the \emph{total} energy and so $N$
is the total number of nodes and not the total number of dilations
$N_{\rm I}$. In constrast, the staggered susceptibility was extracted
just from the up/down fluctuations of the buckled sites. From finite-size
scaling we have $C\propto |T-T_c|^{-\alpha}\propto L^{\alpha/\nu}$. In
Fig.~\ref{fig:C-T_and_CMax-L}, we fit the peaks of $C$ with a
power law function and find $\alpha/\nu=0.068\pm0.018$ for positive dilations. For negative
dilations (see Fig.~\ref{fig:C-T_and_CMax-Lneg}), we find
$\alpha/\nu=0.074\pm0.016$. Note that these values are smaller than
the 3D Ising class ($n=1$ spin component) $\alpha/\nu\simeq0.175$ and
larger than the 2D Ising class $\alpha/\nu=0({\rm log})$. Although we
can also fit our data equally well with a logarithmic function, we
note that these estimates are $\sim3.5$ and $\sim4.5$ standard
deviations away from $\alpha=0$, consistent with the possibility of a different
universality class for this highly compressible Ising model.  
% We also want to note that
%we have a membrane (manifold) with a dimensionality equals to 2
%embeded in a 3 dimensional space.
%We also extract the exponent $\alpha$ by fitting $C$ with a power law
%function. Note that the value of exponent will be sensitive to the
%choice of $T_c$. We will use $T_c(L\rightarrow \infty)$ obtained from
%fitting to extract $\nu$.

\subsection{Critical temperature $T_c$ and correlation length exponent $\nu$}
\begin{figure}[htp]
\begin{center}
\includegraphics[width=0.8\textwidth]{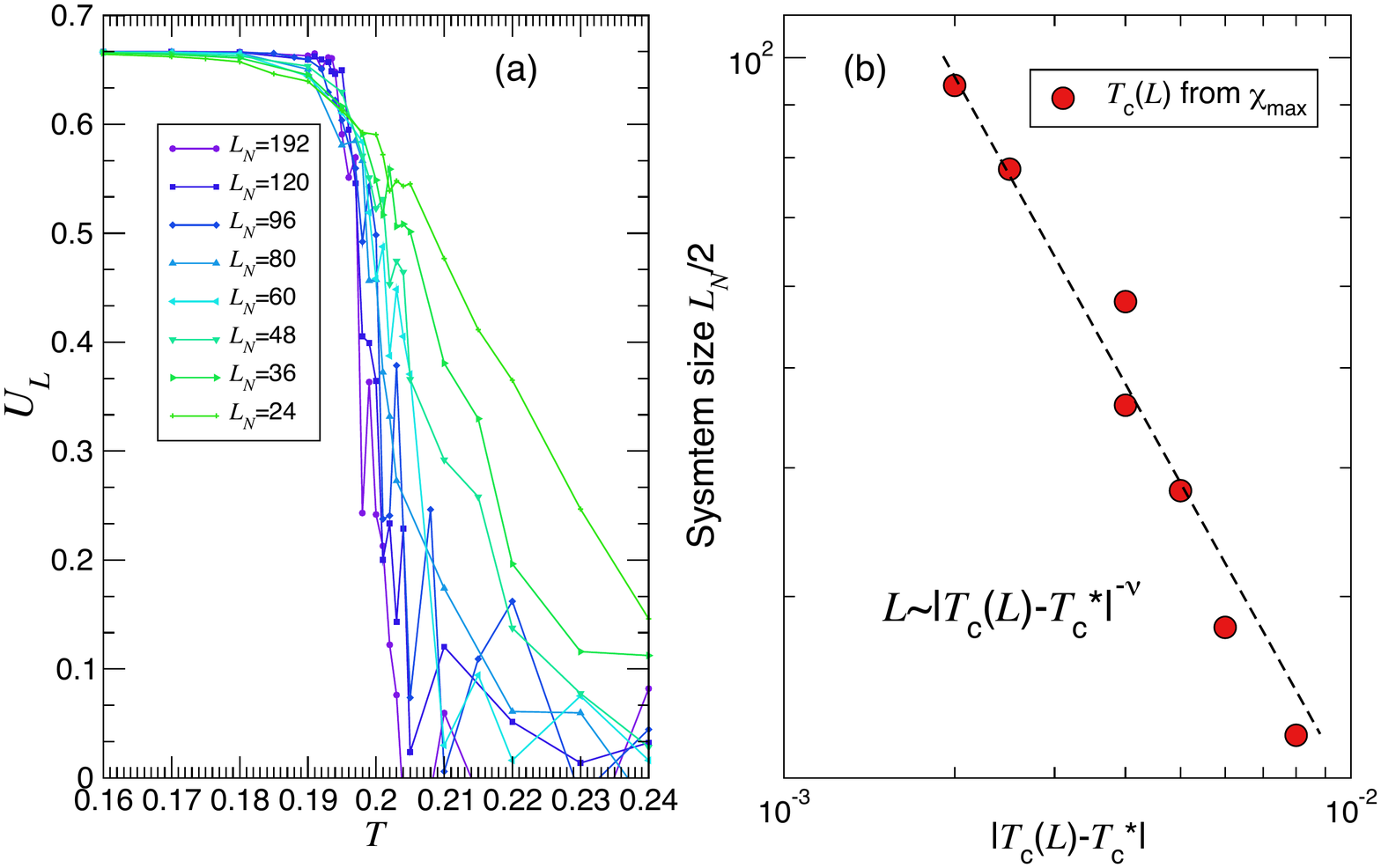}
\end{center}
\caption{(a) Fourth order Binder cumulant for $\Omega_0>0$ membranes as a function of $T$ for
  different linear system sizes $L_N$.  Data for $L_N=160$ are not
  shown for clarity. $U_L$ approaches 2/3 at low $T$ ($T<T_c$) and
  goes to 0 at high $T$ ($T>T_c$). (b) Log-log plot of linear system
  size $L_N/2$ as a function of $|T_c(L)-T_c^*|$. $T_c(L)$ are
  obtained from the peaks of $\chi'$ and $T_c^*$ is the estimated
  $T_c(\infty)$ obtained from the crossing of $U_L$ of the three largest
  systems. The data are fitted with a power law function and we find
  $\nu=0.99\pm0.10$.}
\label{fig:UL-T_and_Tc}
\end{figure}
\begin{figure}[htp]
\begin{center}
\includegraphics[width=0.8\textwidth]{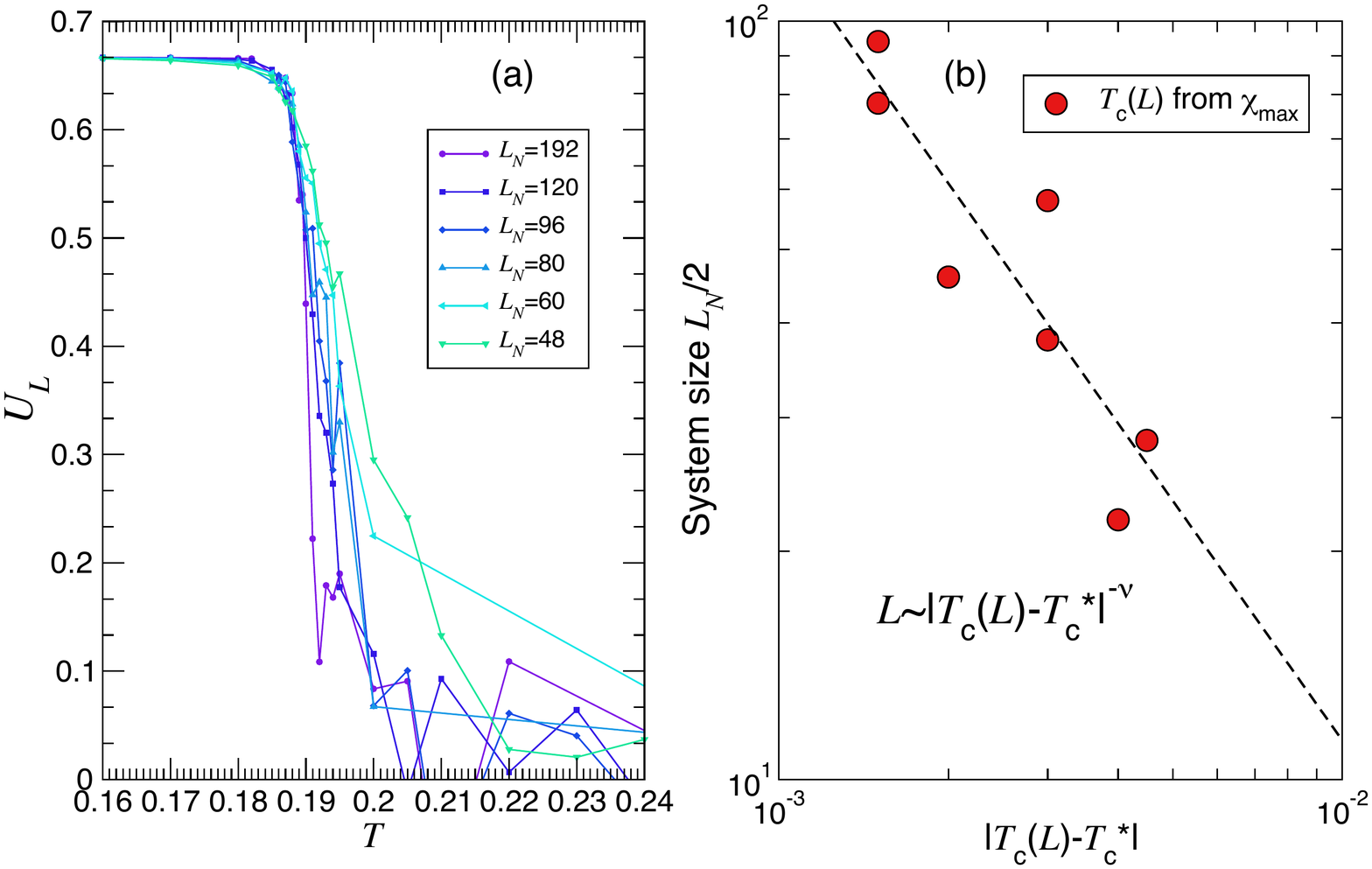}
\end{center}
\caption{(a) Fourth order Binder cumulant for $\Omega_0<0$ membranes as a function of $T$ for
  different linear system sizes $L_N$.  Data for $L_N=160$ are not
  shown for clarity. (b) Log-log plot of linear system size $L_N/2$ as
  a function of $|T_c(L)-T_c^*|$. $T_c(L)$ are obtained from the peaks
  of $\chi'$ and $T_c^*$ is the estimated $T_c(\infty)$ obtained from
  the crossing of $U_L$ of the three largest systems. The data points are
  fitted with a power law function and we find $\nu=1.05\pm0.25$.}
\label{fig:UL-T_and_Tc_stitches}
\end{figure}
The simplest way to estimate the correlation length exponent $\nu$ is
to assume that, as in virtually all known critical phenomena,
hyperscaling ($d \nu=2 -\alpha$) holds, which in $d=2$ dimensions
gives $\nu=\frac{1}{1+ \alpha/(2\nu)}$. Thus, from our specific heat
data, we find $\nu=0.967 \pm 0.008$ for positive dilations and
$\nu=0.964 \pm 0.007$ for negative dilations, results which are
approximately 4-5 standard deviations away from the incompressible
Ising value $\nu=1$.

As a consistency check, we can also extract
$\nu$ using finite size scaling by fitting
$L=c_0|T_c(L)-T_c(\infty)|^{-\nu}$, where $T_c(L)$ are obtained from
the peak location of susceptibility, $T_c(\infty)$ is the critical
temperature in the thermodynamic limit, and $c_0$ is a
constant. Extracting $\nu$ by this method is rather difficult because we have
\emph{three} adjustable parameters and we do not know the true $T_c$
in the thermodynamic limit.  We use the fourth order Binder
cumulant~\cite{binder-RepProgPhys-1997, sandvik-AIP-2010}
\begin{equation}
U_L=1-\frac{\langle\mst^4\rangle}{3\langle\mst^2\rangle^2},
\end{equation}
to locate the phase transition and to estimate $T_c(\infty)$. $U_L$
approaches 2/3 in the low temperature broken symmetry phase and
approaches 0 in the symmetric phase. $T_c$ is usually found by
finding a $T$ where the two curves $U_L$ cross. Unfortunately, our
data, for this highly compressible Ising model, are quite noisy, and we do
not observe systematic dependence between intersection points and
system size. We thus use the crossing of the three largest systems to
estimate $T_c^*=T_c^{\rm estimated}(\infty)\simeq0.194$ for $\Omega_0>0$. With this value of $T_c^*$, we fit our data with a power law and
find $\nu=0.99\pm0.10$, shown in Fig.~\ref{fig:UL-T_and_Tc}.

For membranes with $\Omega_0<0$ we find
$T_c^*=T_c^{\rm estimated}(\infty)\simeq0.188$ and
$\nu=1.05\pm0.25$. Data are shown in
Fig.~\ref{fig:UL-T_and_Tc_stitches}. Note that while consistent with a
value of $\nu$ near one, the error bars are much bigger than those
obtained by using the formula $\nu=\frac{1}{1+\alpha/(2\nu)}$.
%Because
%the extracted $\nu$ is sensitive to $T_c(\infty)$, we also fitted our
%data with $T_c^*=0.1945$ and $T_c^*=0.195$ and we obtained
%$\nu=0.83\pm0.09$ and $\nu=0.66\pm0.08$, respectively. We will compare
%these values of $\nu$ with $\nu$ obtained from the hyper scaling
%relation $\alpha=2-d\nu$ and the extracted $\alpha/\nu$ from $C$. We
%can write $\nu=2/(d+\frac{\alpha}{\nu})$. Using $\alpha/\nu=0.11$
%obtained from simulations, we obtain $\nu(d=2)\simeq0.95$ and
%$\nu(d=3)\simeq0.64$. Two different finite size scaling measurements,
%from $C$ and $T_c$, suggest that $\nu\simeq1$.
%With our limited simulations and not knowing the
%true value of ciritcal temeperature in the thermodynamic limit, it is
%hard to make a conclusive value of $\nu$.
\subsection{Magnetization $\mst$ and critical exponent $\beta$}\label{critbeta}
\begin{figure}[htp]
\begin{center}
\includegraphics[width=0.8\textwidth]{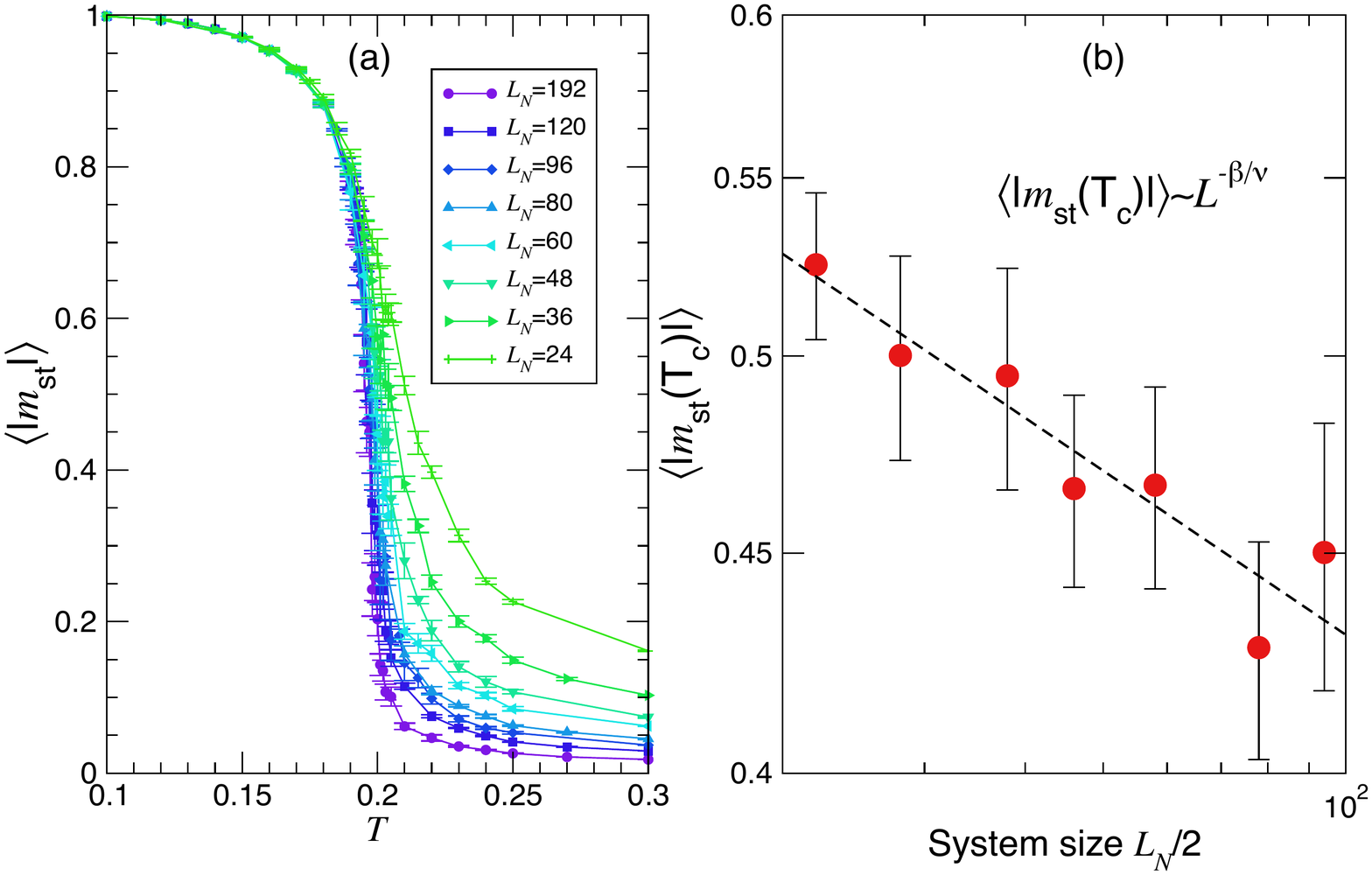}
\end{center}
\caption{(a) Absolute staggered magnetization $\langle |\mst|\rangle$ for $\Omega_0>0$ as
  a function of temperature $T$ for different system sizes. Data for
  $L_N=160$ are not shown for clarity.  Log-log plot of $|\mst|$ at
  $T_c(L)$ as a function of linear system size $L_N/2$. The data with
  $L_N/2>20$ is fitted with a power law function and we find
  $\beta/\nu=0.127\pm0.038$.}
\label{fig:mst-T_and_mst-L}
\end{figure}
\begin{figure}[htp]
\begin{center}
\includegraphics[width=0.8\textwidth]{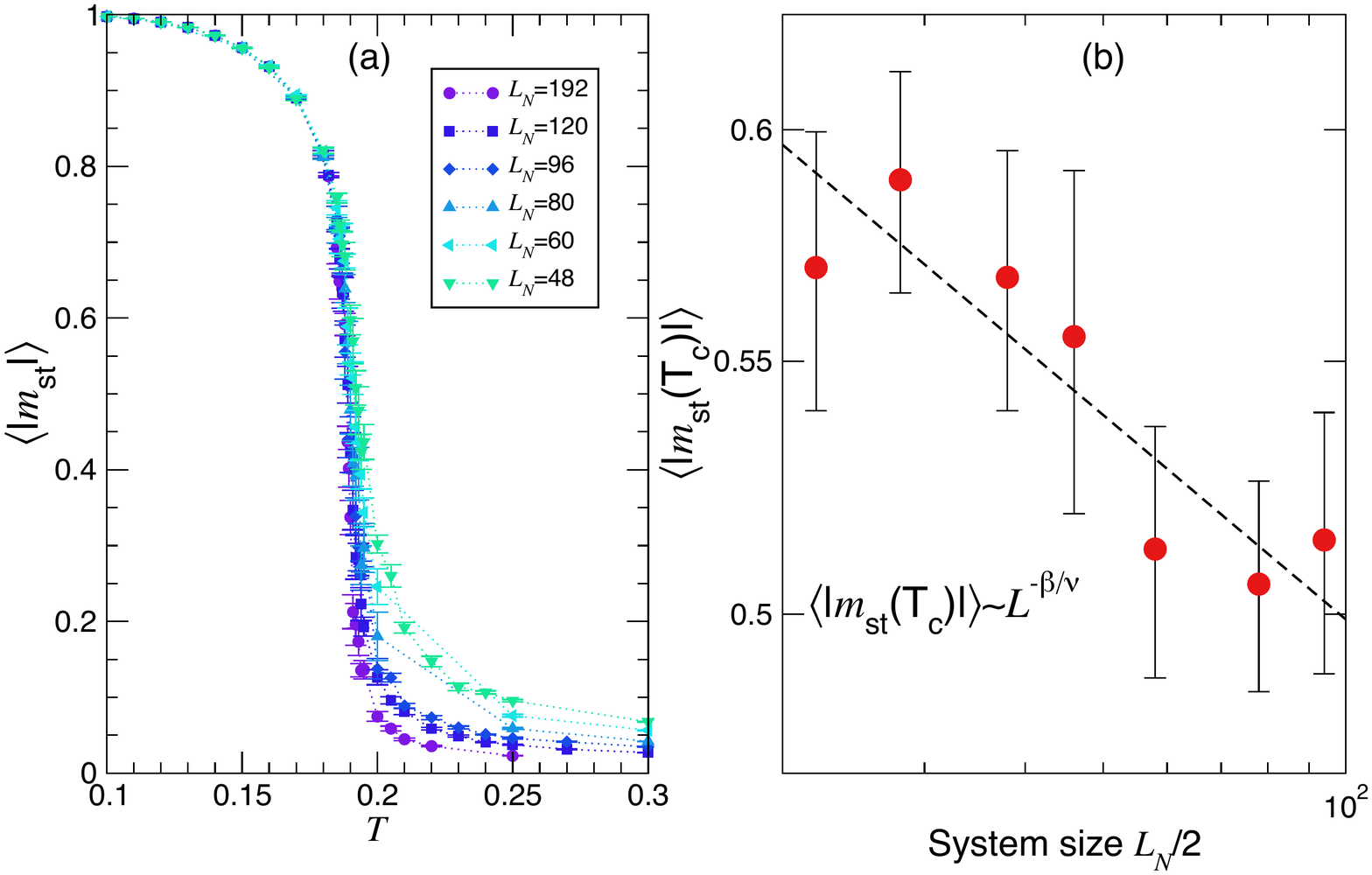}
\end{center}
\caption{(a) Absolute staggered magnetization $\langle |\mst|\rangle$ for $\Omega_0<0$ as
  a function of $T$ for different system sizes. Data for $L_N=160$ are
  not shown for clarity. Log-log plot of $|\mst|$ at $T_c(L)$ as a
  function of linear system size $L_N/2$. The data points are fitted
  with a power law function and we find $\beta/\nu=0.111\pm0.036$.}
\label{fig:mst-T_and_mst-L_sticthes}
\end{figure}
The quantity $|\mst|$ as a function of $T$ for $\Omega_0>0$ is shown in
Fig.~\ref{fig:mst-T_and_mst-L}(a). One way to extract the staggered
magnetization exponent $\beta$ is to fit $|\mst|$ at $T=T_c(L)$ with
$L^{-\beta/\nu}$. In our system, it is hard to get reliable values of
$\sqrt{\mst^2}$ at $T=T_c(L)$ with the error bars and the temperature
steps we have in our simulations. Hence we obtain
$\beta/\nu=0.127\pm0.038$ with a large uncertainty. Data for negative
dilations are shown in Fig.~\ref{fig:mst-T_and_mst-L_sticthes} and we
find $\beta/\nu=0.111\pm0.036$.
\section{Comparison of fluctuations in empirical graphene model and membrane model}
The scaling behavior of critical systems, such as freestanding
thermalized membranes, is known to be insensitive to microscopic
details \cite{nelsonbook, katsnelson2012graphene}. To demonstrate this
explicitly, we performed computationally costly molecular dynamics
simulations of graphene with AIREBO
potentials~\cite{stuart2000reactive} on Large-scale Atomic/Molecular
Massively Parallel Simulator (LAMMPS)~\cite{plimptonLAMMPS} and
measured the height-height correlations $|h({\bf q})|^2$.  Details on
molecular dynamics procedures to simulate graphene with AIREBO
potentials using LAMMPS can found in
~\cite{hanakata-PRL-121-255304-2018, hanakata-PRR-2-042006-2020}. The
height fluctuations in Fourier space $\langle |h({\bf q})|^2\rangle$
at zero strain are given by
\begin{equation}
  \langle |h({\bf q})|^2\rangle=\frac{k_{\rm B}T}{A\kappa(q) q^4},
  \label{eq:harmonic}
\end{equation}
where $A$ is the area of the membrane, $\kappa(q)$ is the renormalized
bending rigidity, and $q$ is magnitude of the wave vector. The
effective $\kappa$ and $Y$ as a function of wavevector are given
by~\cite{kosmrlj-PRB-93-12-125431, bowick-PRB-95-104109-2017}
\begin{align}
    \kappa(q) &\propto
                \begin{cases}
                  \kappa_0     & \quad \text{if } q\gg q_{\rm th}\\
                  \kappa_0\left(\frac{q}{q_{\rm th}}\right)^{-\eta}  & \quad \text{if } q\ll q_{\rm th}
                \end{cases}\\
    Y(q) &\propto
           \begin{cases}
             Y_0     & \quad \text{if } q\gg q_{\rm th}\\
             Y_0\left(\frac{q}{q_{\rm th}}\right)^{\eta_u}  & \quad \text{if } q\ll q_{\rm th}, 
           \end{cases}
  \end{align}
  where $\eta\sim0.8-0.85$ and $\eta_u\sim0.3-0.4$. The thermal length is given by 
  \begin{equation}
  l_{\rm th}=\sqrt{\frac{\pi^316\kappa^2_0}{3k_{\rm B}T\,Y_0}}\equiv \frac{\pi}{q_{\rm th}}, 
\end{equation}
where $T$ is the temperature, $k_{\rm B}$ is the Boltzmann constant,
$\kappa_0$ is the bending rigidity at $T=0$, and $Y_0$ is the 2D
Young's modulus at $T=0$. Thus we expect
$\langle |h({\bf q})|^2\rangle\propto q^{-4+\eta}$ for the temperature
range we are interested in. Stiffening occurs when the observed length
scale becomes comparable to the thermal length
$l_{\rm th}$~\cite{kosmrlj-PRB-93-12-125431}.
  
\begin{figure}
  \centering
  \includegraphics[width=0.8\textwidth]{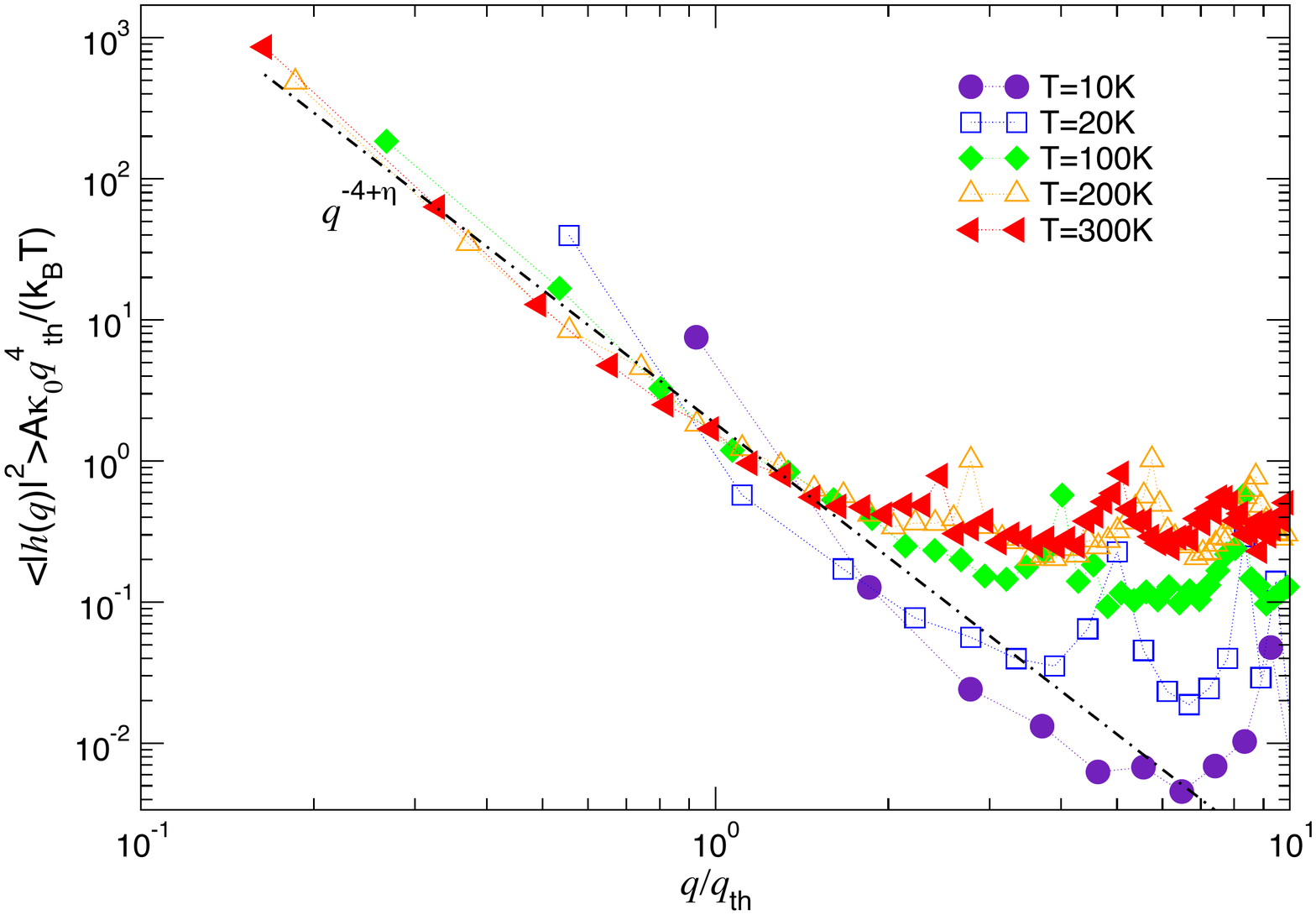}
  \caption{The collapse of normalized $\langle |h(q)|^2\rangle$ as a
    function of $q/q_{\rm th}$ for graphene modeled with an AIREBO
    potential simulated at $T=[10, 20, 100, 300]$~K.  The membrane has
    a length of $L=200~\AA$ and a width of $W=50~\AA$. Data are
    obtained by taking 100 snapshots during a total run of 10ns
    ($10^7$ time steps). The time interval between snapshots is
    100~ps. The dotted dashed line shows the $q^{-(4-\eta)}$
    behavior.}
  \label{fig:h2q-graphene}
\end{figure}

For graphene with an AIREBO potential, $\kappa_0=1$~eV and
$Y_0=21$~eV/\AA$^2$---thus, $l_{\rm th}\sim 20$~\AA~at room
temperature ($T=300$~K). We indeed see a collapse of the rescaled
$|h({\bf q})|^2$ in Fig. \ref{fig:h2q-graphene} when the data are
plotted using Eq. \ref{eq:harmonic}. (see also
Ref.~\cite{bowick-PRB-95-104109-2017}).

\begin{figure}
  \centering
  \includegraphics[width=0.8\textwidth]{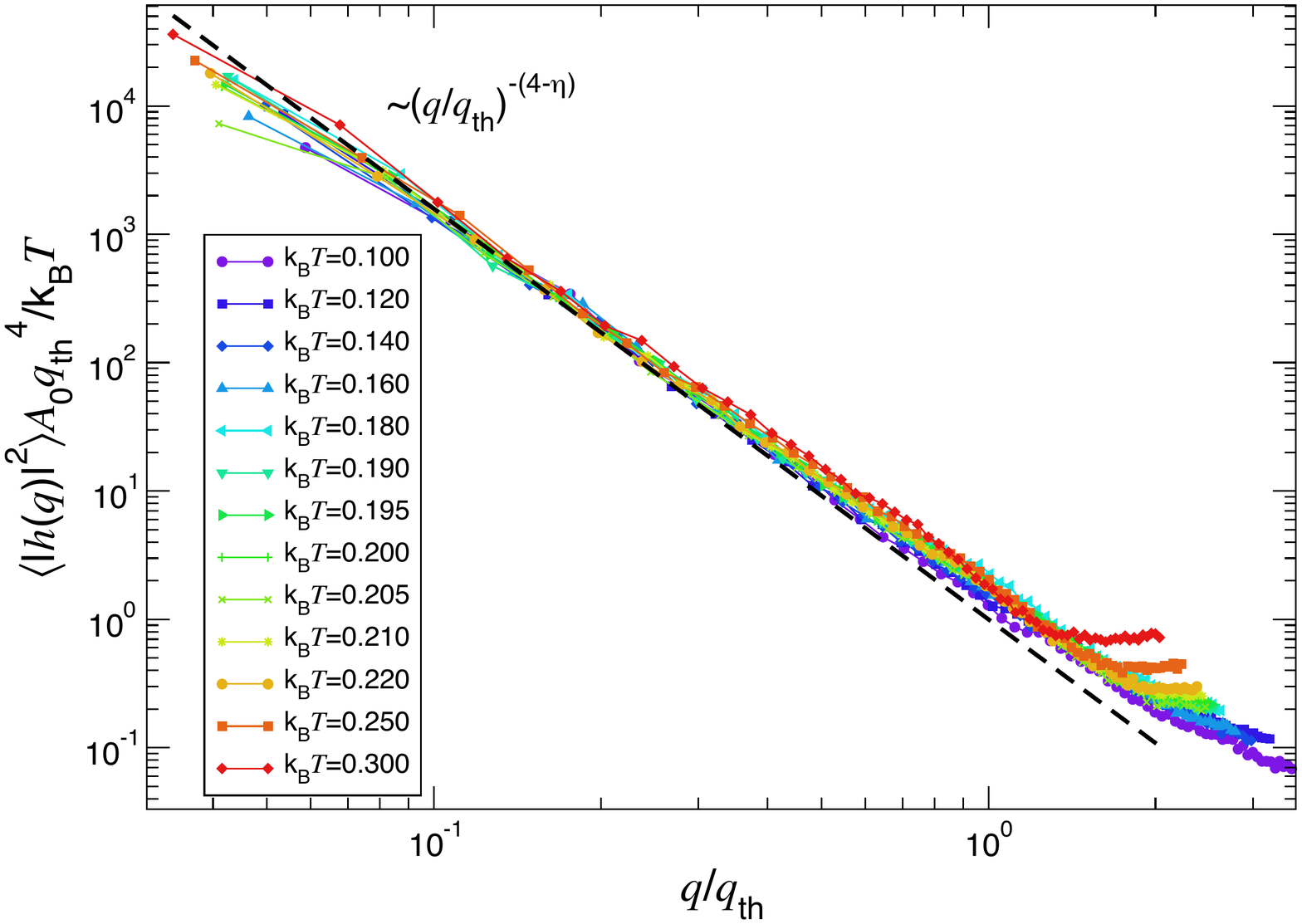}
  \caption{Normalized $\langle |h(q)|^2\rangle$ for a square lattice
    membrane of size 120$a_0$x120$a_0$ as a function of $q/q_{\rm th}$
    at different temperatures. The dashed line shows the
    $q^{-(4-\eta)}$ behavior.}
  \label{fig:h2q-cg}
\end{figure}

Next, we plot $\langle |h({\bf q})|^2 \rangle $ for a pristine
membrane with the square lattice coarse-grained model used in the
current work, shown in Fig.~\ref{fig:h2q-cg}. We see that both models
show $\langle |h({\bf q})|^2\rangle $ with $q^{-(4-\eta)}$ behavior
with $q\simeq0.8$.

%\bibliography{biblio}